\documentclass[oldversion]{aa}
\usepackage{txfonts}
\usepackage{natbib}
\usepackage{subfigure}
\bibpunct{(}{)}{;}{a}{}{,} 
\usepackage{graphicx}
\usepackage{amssymb}
\usepackage{threeparttable}
\usepackage{rotating}
\usepackage{epsfig}

\usepackage{epstopdf}

\newcommand{\Msun}{\mathrm{M}_{\odot}}
\newcommand{\mdot}{\mathrm{M}_{\odot}~\mathrm{yr}^{-1}}
\newcommand{\lum}{\mathrm{erg~s}^{-1}}
\newcommand{\fluence}{\mathrm{erg~cm}^{-2}}
\newcommand{\flux}{\mathrm{erg~cm}^{-2}~\mathrm{s}^{-1}}
\newcommand{\cnts}{\mathrm{counts~s}^{-1}}

\newcommand{\ascabron}{AX~J1745.6--2901}
\newcommand{\ascabroncxo}{CXOGC~J174535.6--290133}
\newcommand{\xmmbron}{XMM~J174457--2850.3}
\newcommand{\grsbron}{GRS~1741--2853}
\newcommand{\brontwee}{CXOGC~J174535.5--290124}
\newcommand{\brondrie}{CXOGC~J174540.0--290005}
\newcommand{\bronvier}{Swift~J174553.7--290347}
\newcommand{\bronviercxo}{CXOGC~J174553.8--290346}
\newcommand{\bronvijf}{Swift~J174622.1--290634}
\newcommand{\bronacht}{CXOGC~J174538.0--290022}
\newcommand{\sgra}{Sgr~A$^{*}$}
\newcommand{\adcbron}{CXOGC~174540.0--290031}
\newcommand{\munotransient}{CXOGC~174554.3--285454}
\newcommand{\xmmbrontwee}{XMM~J174544--2913.0}

\newcommand{\swift}{\textit{Swift}}
\newcommand{\chan}{\textit{Chandra}}
\newcommand{\xmm}{\textit{XMM-Newton}}
\newcommand{\inte}{\textit{Integral}}

\newcommand{\rxte}{\textit{RXTE}}

\begin{document}

\title{A four-year baseline \swift\ study of enigmatic X-ray transients located near the Galactic center}

\titlerunning{Swift monitoring of enigmatic X-ray transients}

\author{Nathalie Degenaar\thanks{e-mail: degenaar@uva.nl} \and Rudy Wijnands}

\authorrunning{N. Degenaar \and R. Wijnands}

\institute{University of Amsterdam, Postbus 94249, 1098 SJ, Amsterdam, the Netherlands}

\date{Received 2 July 2010 / Accepted 21 September 2010}

\abstract {
We report on continued monitoring observations of the Galactic center carried out by the X-ray telescope aboard the \swift\ satellite in 2008 and 2009. This campaign revealed activity of the five known X-ray transients \ascabron, \brontwee, \grsbron, \xmmbron\ and \bronacht. All these sources are known to undergo very faint X-ray outbursts with 2--10 keV peak luminosities of $L_{\mathrm{X, peak}}\sim10^{34-36}~\lum$, although the two confirmed neutron star low-mass X-ray binaries \ascabron\ and \grsbron\ can also become brighter ($L_{\mathrm{X, peak}}\sim10^{36-37}~\lum$).
We discuss the observed long-term lightcurves and X-ray spectra of these five enigmatic transients. In 2008, \ascabron\ returned to quiescence following an unusually long accretion outburst of more than 1.5~years. \grsbron\ was active in 2009 and displayed the brightest outburst ever recorded for this source, reaching up to a 2--10 keV luminosity of $L_{\mathrm{X}}\sim 1 \times 10^{37}~(D/7.2~\mathrm{kpc})^{2}~\lum$. This system appears to undergo recurrent accretion outbursts approximately every 2~years. Furthermore, we find that the unclassified transient \xmmbron\ becomes bright only during short episodes (days) and is often found active in between quiescence ($L_{\mathrm{X}}\sim10^{32}~\lum$) and its maximum outburst luminosity of $L_{\mathrm{X}}\sim10^{36}~\lum$. \brontwee\ and \bronacht, as well as three other very-faint X-ray transients that were detected by \swift\ monitoring observations in 2006, have very low time-averaged mass-accretion rates of $\langle \dot{M} \rangle _{\mathrm{long}}\lesssim 2 \times 10^{-12}~\mdot$. Despite having obtained two years of new data in 2008 and 2009, no new X-ray transients were detected. 
}

\keywords{X-rays: binaries - Stars: neutron - Accretion, accretion disks - Galaxy: center - X-rays: individuals: \ascabron, \brontwee, \grsbron, \xmmbron, \bronacht}

\maketitle 


\section{Introduction}\label{sec:intro}
Starting in 2006 February, the \swift\ satellite has been monitoring the Galactic center (GC) with the onboard X-ray telescope \citep[XRT;][]{burrows05}. In this campaign, short ($\sim$ 1 ks) pointings are carried out on an almost daily basis\footnote{Except during the months November--February, when the GC is too close (within 45 degrees) to the Sun.}, covering a field of $\sim26'\times26'$ of sky around \sgra\ \citep{kennea_monit,degenaar09_gc}. This is an ideal setting for detecting transient X-ray sources in one of the most active X-ray regions in the Milky Way. 

X-ray transients alternate periods of quiescence, that have a typical duration of years to decades and are characterized by 2--10 keV luminosities of $L_{X} \sim 10^{30-33}~\lum$, with occasional outbursts during which the X-ray luminosity increases by a factor $\gtrsim100$ for weeks to months. A large fraction of the galactic X-ray transients can be identified with neutron stars or black holes accreting matter from a companion star in an X-ray binary. Based on the nature of the donor star, we can distinguish low-mass X-ray binaries (LMXBs; $M_{\mathrm{donor}} \lesssim 1~\Msun$, spectral type later than B) or high-mass X-ray binaries (HMXBs; $M_{\mathrm{donor}} \gtrsim 10~\Msun$, spectral type O or B).

In LMXBs, matter is generally transferred because the donor star fills its Roche lobe, a process that involves the formation of an accretion disk. In such systems, the transient behavior is explained in terms of a thermal-viscous instability that causes the disk to oscillate between a cold, neutral state (quiescence), and one in which it is hot and ionized, causing a strong increase in the mass-accretion rate and resulting in an X-ray outburst \citep[e.g.,][]{king98,lasota01}. During quiescence, the disk regains the mass that was lost during the outburst and the cycle repeats. Symbiotic X-ray binaries form a small sub-class of LMXBs  in which the compact primary, most likely a neutron star, is accreting matter from the wind of an M-type giant companion \citep[e.g.,][]{masetti2007}.

Amongst the transient HMXBs, most of the currently known systems are Be/X-ray binaries. In such systems, the compact primary is in a wide and eccentric orbit accreting matter from the circumstellar disk surrounding a main sequence Oe or Be star around periastron passage \citep[e.g.,][]{negueruela04}. However, recently \inte\ and \rxte\ have unveiled a new class of transient HMXBs, called Supergiant Fast X-ray transients \citep[SFXTs; e.g.,][]{negueruela06}, in which the compact star is capturing the strong stellar wind of an O or B supergiant companion. In these systems, the transient behavior is thought to be caused by clumpy or anisotropic winds \citep[e.g.,][]{sidoli09}.

The temporal and spectral properties of the brightest galactic X-ray transients, which have 2--10 keV peak luminosities of $L_{\mathrm{X,peak}} \sim 10^{36-39}~\lum$, are well established through the work of numerous past and present X-ray missions. However, much less is known about transient sources that manifest themselves with lower 2--10 keV peak luminosities of $\sim 10^{34-36}~\lum$ \citep[e.g.,][]{sidoli99,muno05_apj622,porquet05,wijnands06,campana09,heinke09_vfxt}. It is only with the advent of the current generation of sensitive X-ray instruments that the properties of such objects can be studied in detail. 
To date, a few tens of low-luminosity transients are known in our Galaxy. As for the brighter systems, many of these are expected to harbor accreting neutron stars or black holes, but their nature and the underlying mechanism producing their subluminous outbursts is not understood well. 
 
The hypothesis that a significant fraction of the low-luminosity transients are X-ray binaries, gains credence by the detection of thermonuclear X-ray bursts from several of these systems \citep[e.g.,][]{zand1991,maeda1996,cocchi99,cornelisse02, chelovekov07_ascabron,delsanto07,wijnands09}. This establishes that these objects harbor accreting neutron stars, most likely in an LMXB configuration. The observed low X-ray luminosities in combination with estimates of their recurrence times, suggest that these systems have very low time-averaged mass-accretion rates \citep[e.g.,][]{degenaar09_gc}. This might pose a challenge to explain the existence of these LMXBs without having to invoke exotic evolutionary scenarios \citep[e.g.,][]{king_wijn06}. 

Many low-luminosity transients are located in the vicinity of \sgra\ \citep[e.g.,][]{muno05_apj622,wijnands06,degenaar09_gc}. The \swift/GC monitoring program thus provides an excellent setting to detect new low-luminosity transients and to study the long-term behavior of known systems. This opens up the possibility to gain more insight into their duty cycles and the energetics of their outbursts, and thereby to refine estimates of their average mass-accretion rates. This is an important parameter for understanding their evolution \citep[e.g.,][]{king_wijn06}, as well as the properties of thermonuclear X-ray bursts occurring at low accretion luminosities \citep[e.g.,][]{zand05_ucxb,peng2007,cooper07}.

In a previous paper, we discussed a total of seven transients that were found active during \swift/XRT monitoring observations of the GC, carried out in 2006 and 2007 \citep{degenaar09_gc}. Here, we discuss the data accumulated over the years 2008 and 2009, which revealed activity of five previously known X-ray transients.


\section{Observations and data analysis}\label{sec:obs_data}

We obtained all 2008 and 2009 XRT observations of the GC from the \swift~public data archive. For 2008, the nearly daily coverage resulted in a total of 171 observations, amounting to 211 ks of exposure. All data was obtained in the photon counting (PC) mode and covers the epoch 2008 February 19--October 30. In 2009, the campaign was carried out in a slightly different setting, with $\sim 1$~ks observations performed once every $\sim3$~days, instead of the daily repetition between 2006 and 2008. \swift\ targeted the GC from 2009 June 4 till November 1 during 40 observations for a total exposure time of 45 ks. Most of the 2009 data was obtained in the PC mode, apart from a selection of 3 pointings, during which the XRT was operated in the windowed timing (WT) mode. These WT observations were carried out between 2009 October 10--13, and aimed specifically for \grsbron, which was in outburst at that time and caused significant pile-up in the PC data.  

\begin{figure*}
 \begin{center}
          \includegraphics[width=9.0cm]{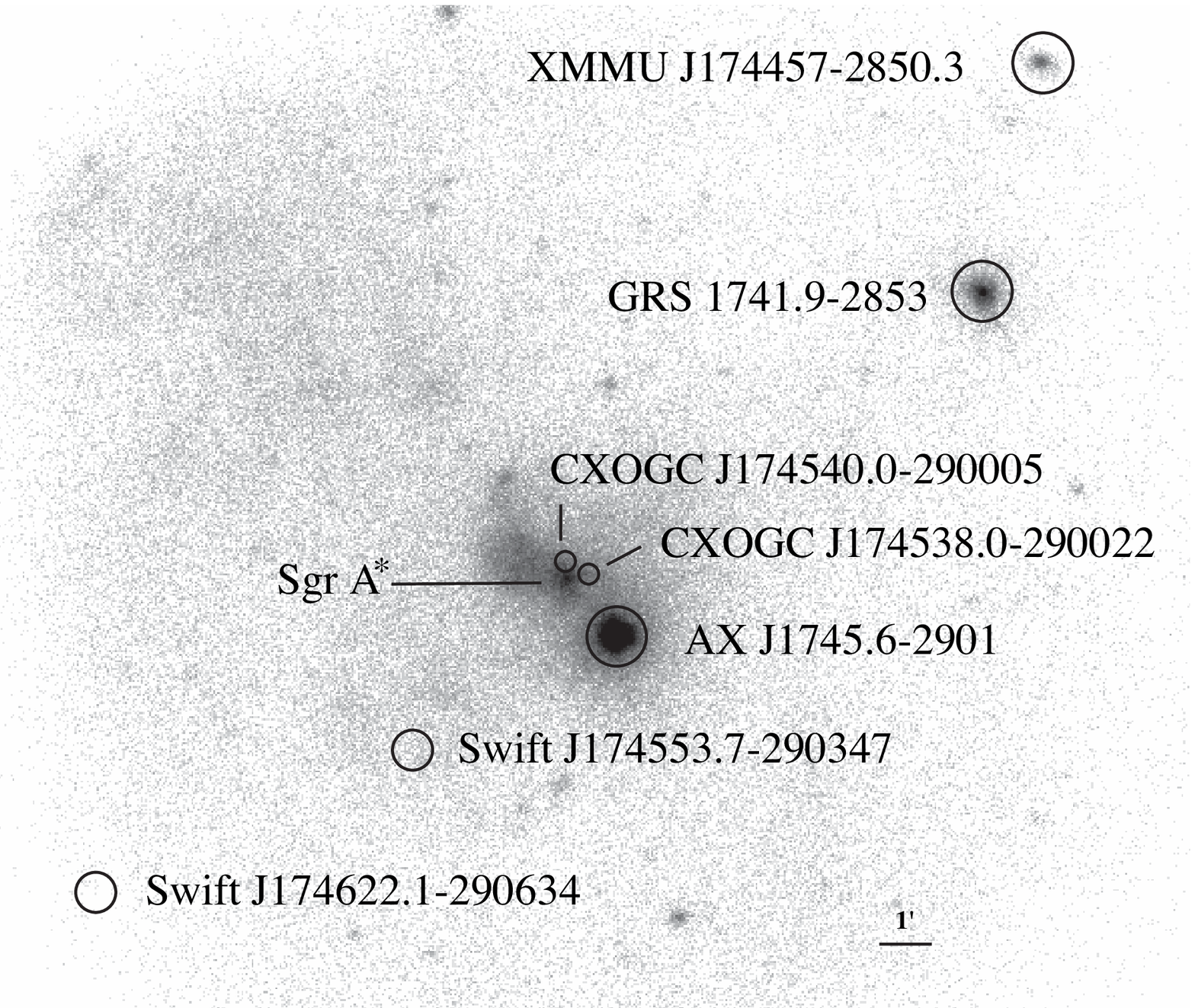}
          \hspace{0.2cm}
    \includegraphics[width=9.0cm]{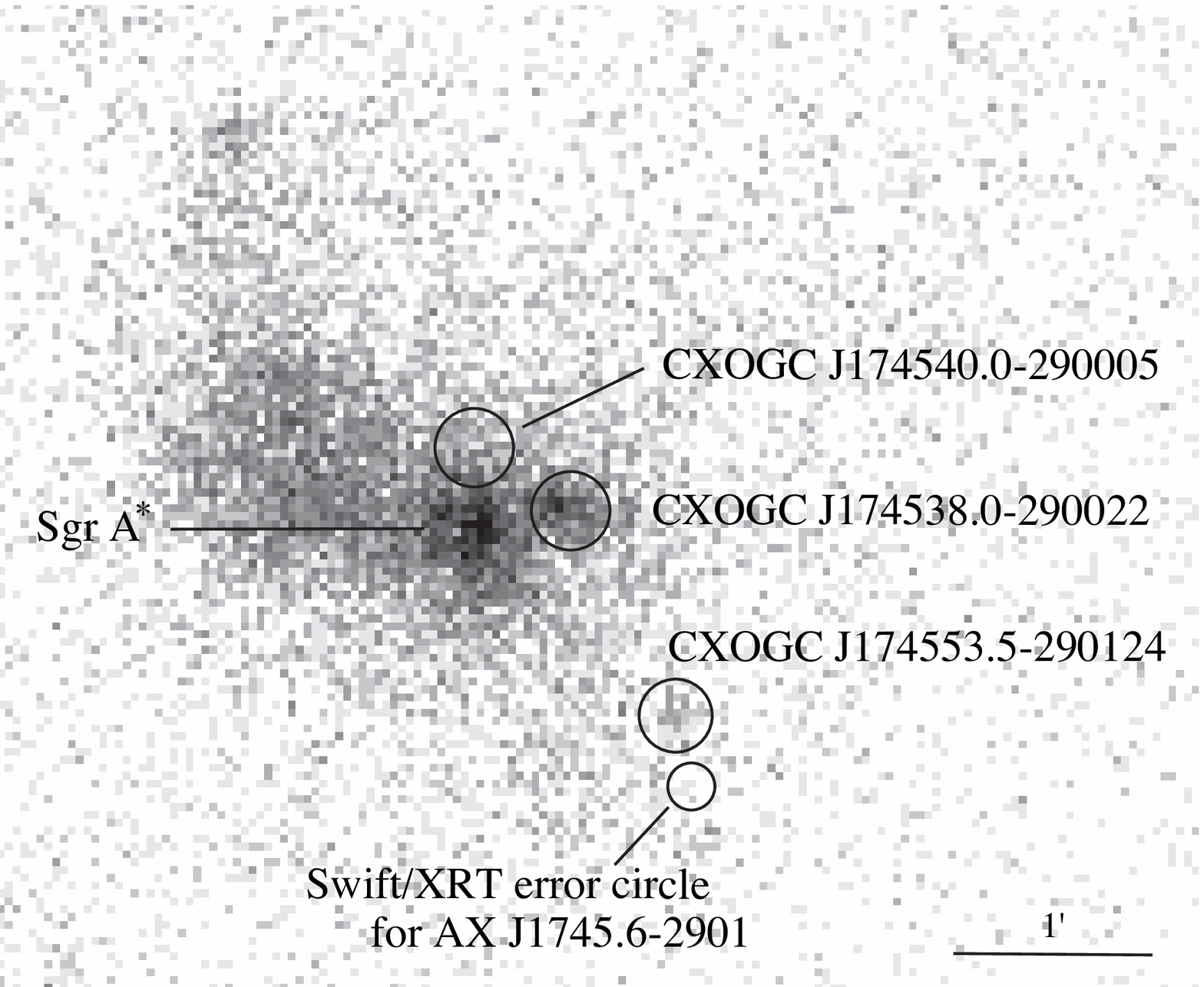}
    \end{center}
\caption[]{\swift/XRT PC mode images (0.3--10 keV) of the region around \sgra. The images indicate the locations of the five X-ray transients that were in outburst in 2008--2009, as well as three additional transients that were detected with \swift\ in 2006 (\brondrie, \bronvier\ and \bronvijf). Left) Merged X-ray image of the data obtained in 2008 and 2009. Right) A magnified image of the inner region around \sgra\ from the epoch between 2008 September 2 and 2009 November 1, during which \ascabron\ resided in quiescence and activity from \brontwee\ could be detected. \bronacht\ was also detected in outburst during that time. }
 \label{fig:ds9}
\end{figure*}

All raw data were processed with the \textsc{xrtpipeline} using standard quality cuts and event grades 0--12 for the PC data and 0--2 for the WT mode observations. Fig.~\ref{fig:ds9} displays two accumulated X-ray images of the \swift/XRT observations. Both images clearly show regions of diffuse emission around \sgra, as well as several X-ray point sources. Five of these can be identified with the known X-ray transients \ascabron, \brontwee, \grsbron, \xmmbron\ and \bronacht, while others are known persistent X-ray sources. In 2006, \swift\ detected three additional X-ray transients in outburst, which are not found active in the new 2008--2009 data set: \brondrie, \bronvier\ and \bronvijf\ \citep[][]{degenaar09_gc}. The locations of these transients are also indicated in Fig.~\ref{fig:ds9}. We used \textsc{Xselect} (v. 2.4) to compare sub-sets of the data to determine when the transients were active. A source was considered in quiescence when it was not detected by visual inspection upon summing multiple observations. A non-detection in 5 ks of data corresponds roughly to a 2--10 keV unabsorbed flux threshold of $\sim 2 \times 10^{-13}~\flux$, or a luminosity of $\sim 2 \times 10^{33}~(D/\mathrm{8~kpc})^2~\lum$, although the exact value depends on the assumed spectral model.

\begin{figure*}
 \begin{center}
    \includegraphics[width=5.4cm]{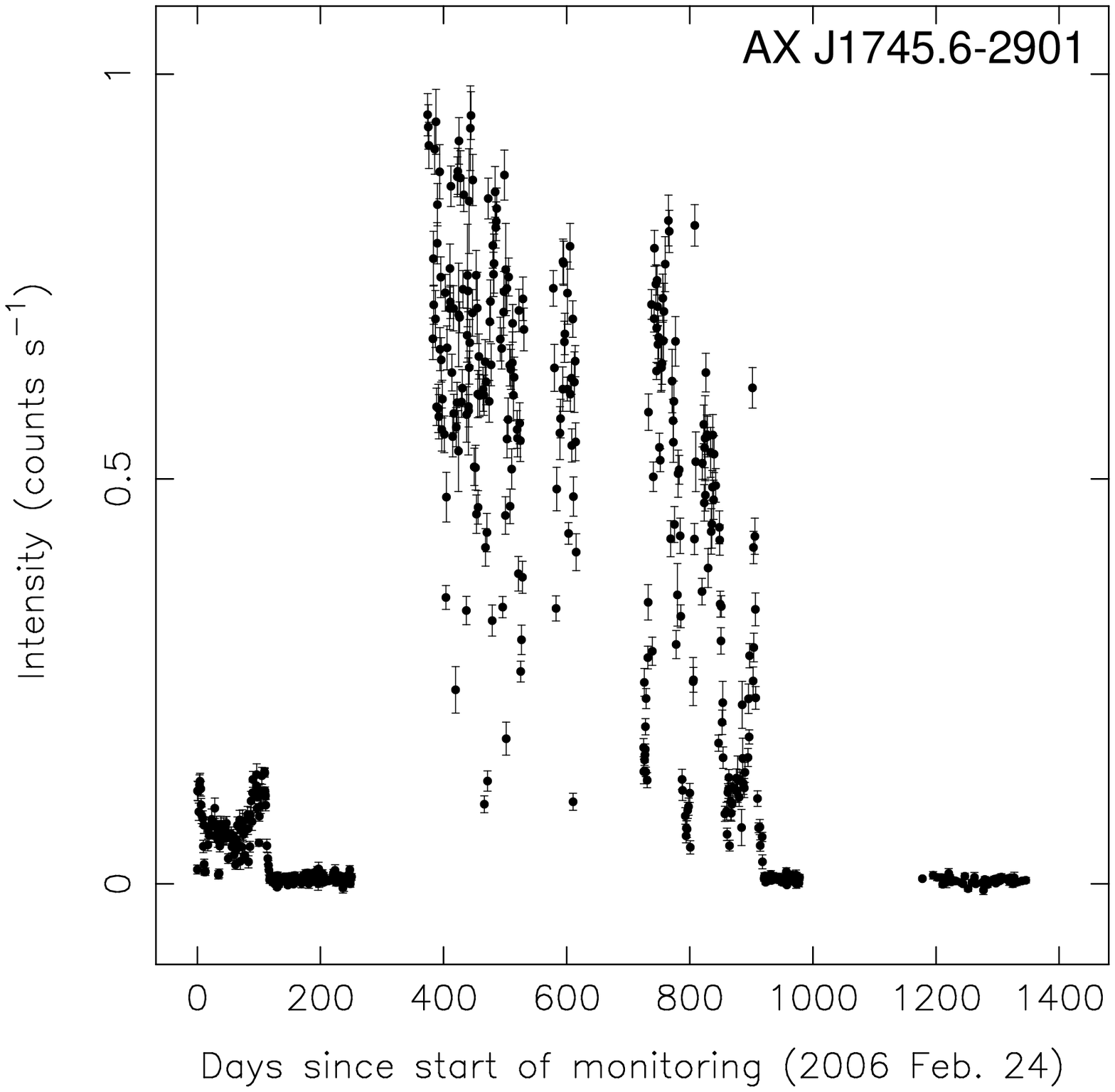}
    \hspace{0.5cm}\vspace{0.3cm}
    \includegraphics[width=5.4cm]{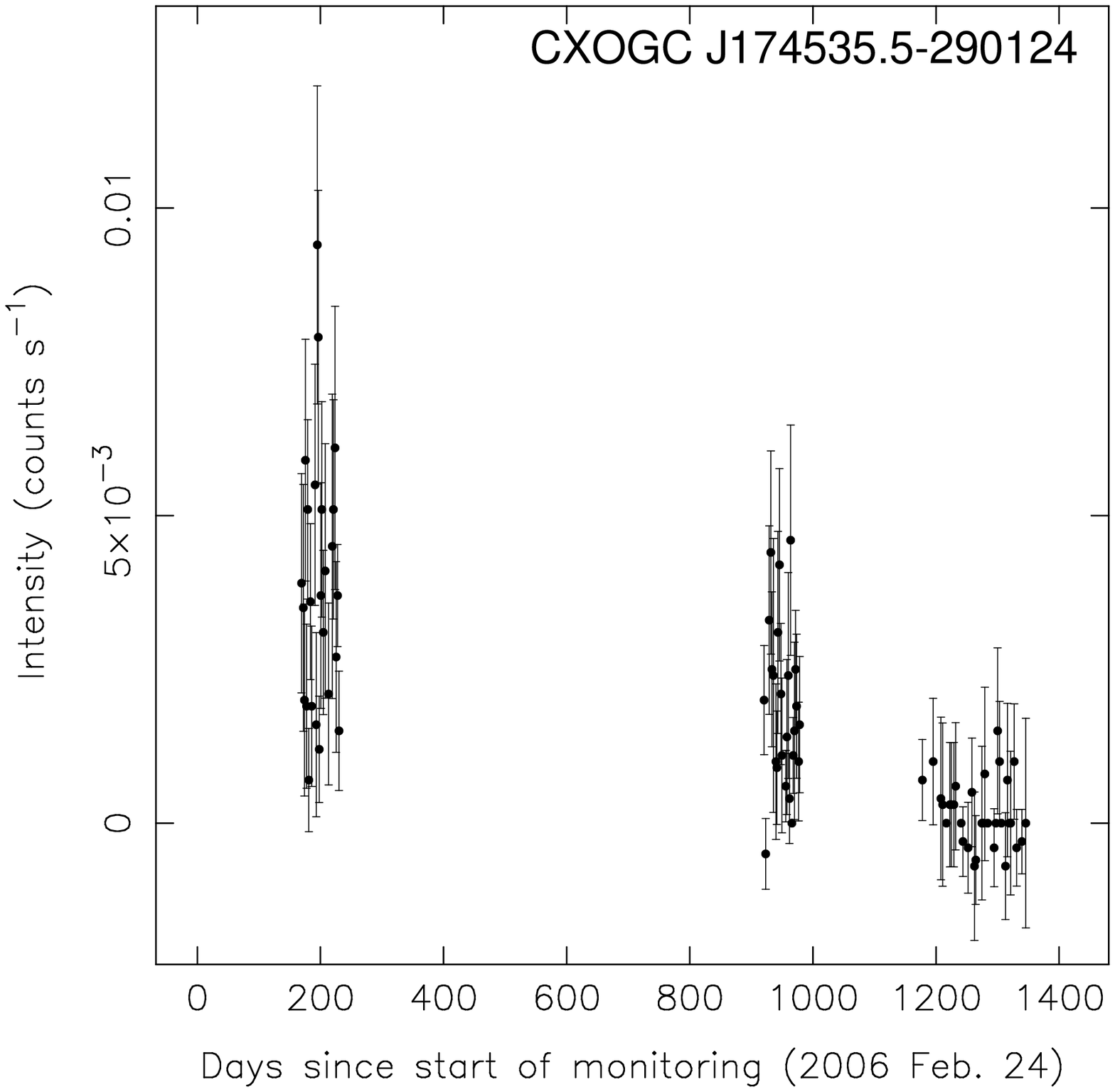}
    \hspace{0.5cm}
         \includegraphics[width=5.4cm]{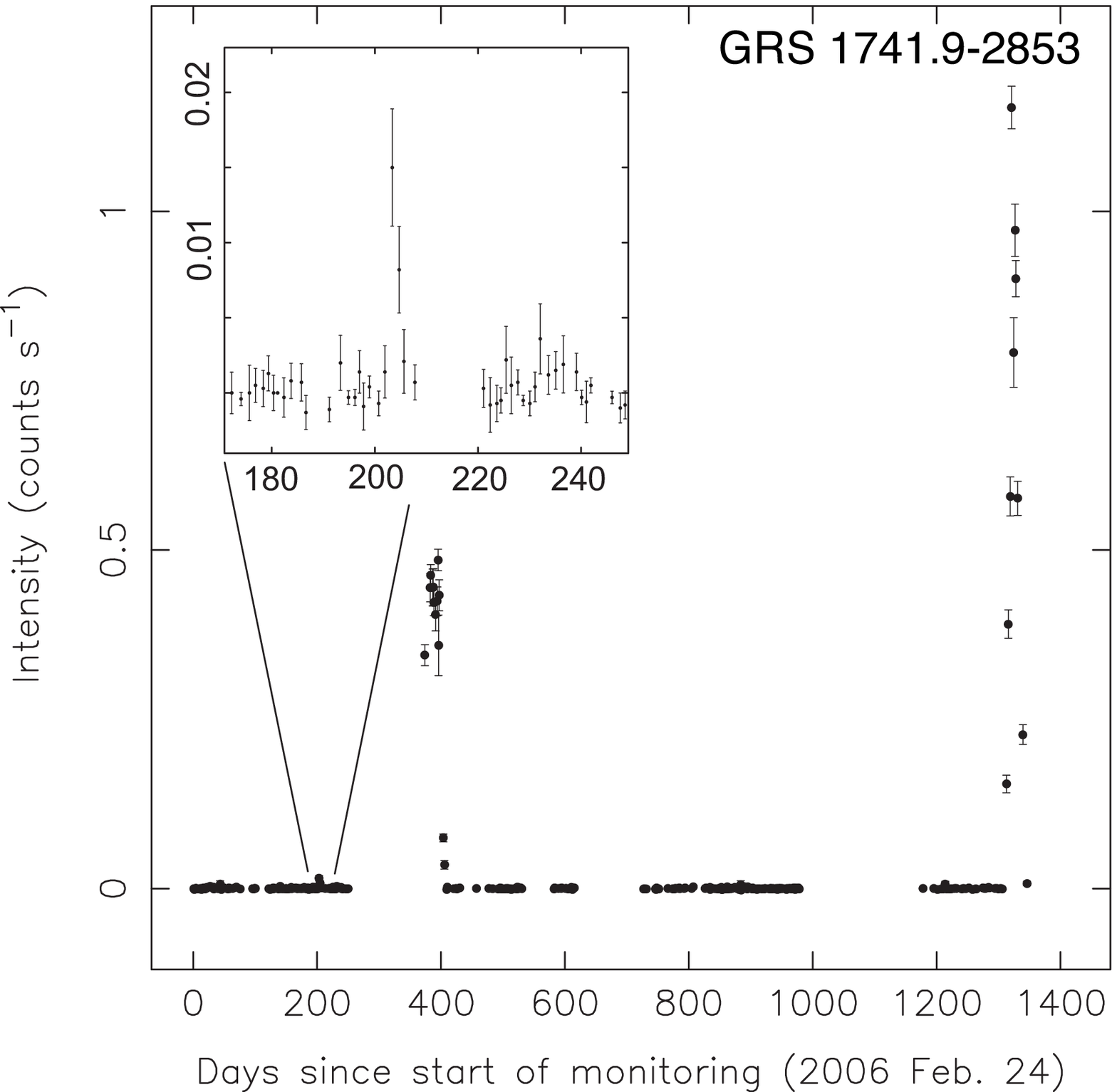}
\hspace{2.5cm}
              \includegraphics[width=5.4cm]{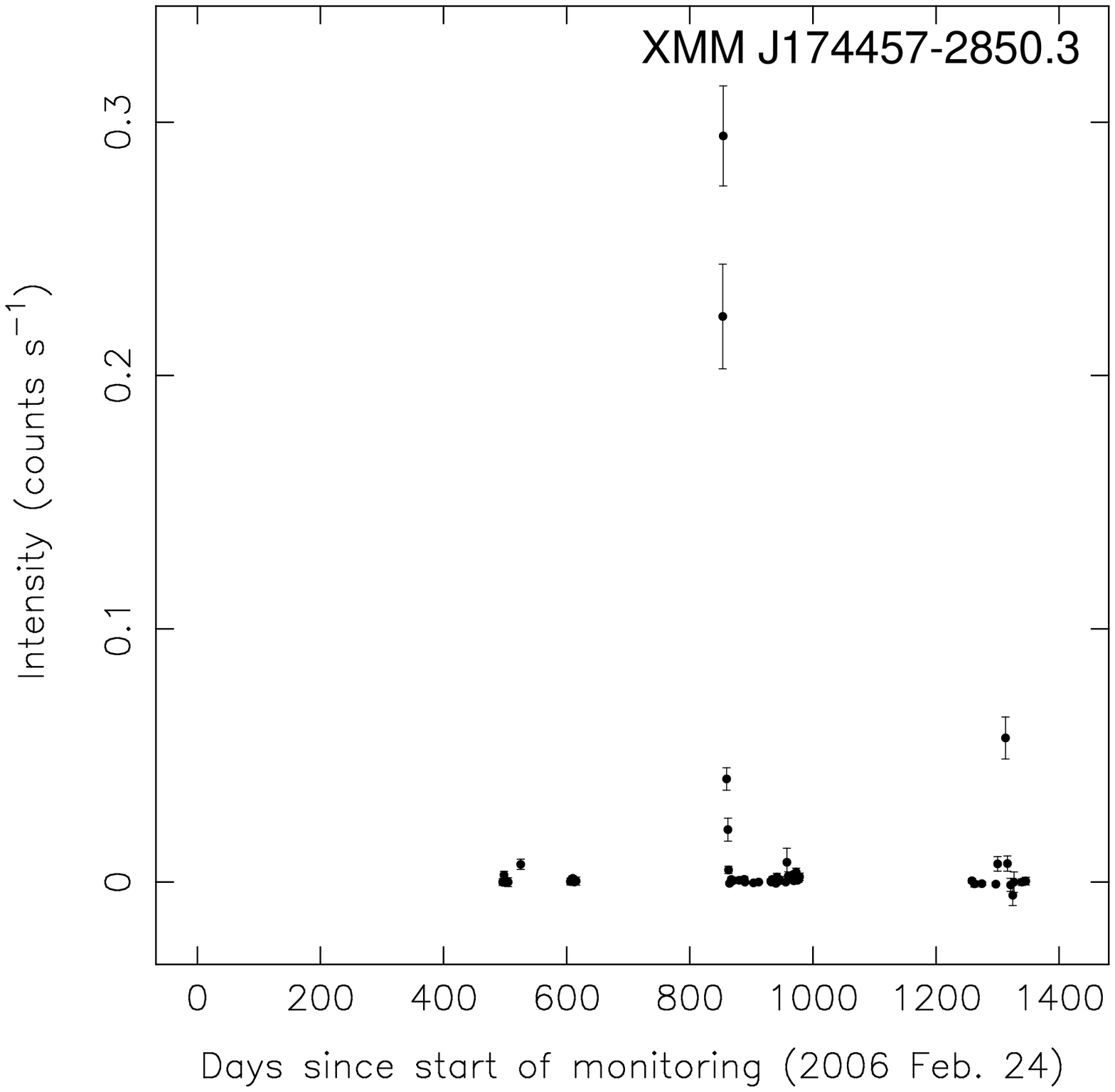}   \hspace{0.5cm}\vspace{0.3cm}
     \includegraphics[width=5.4cm]{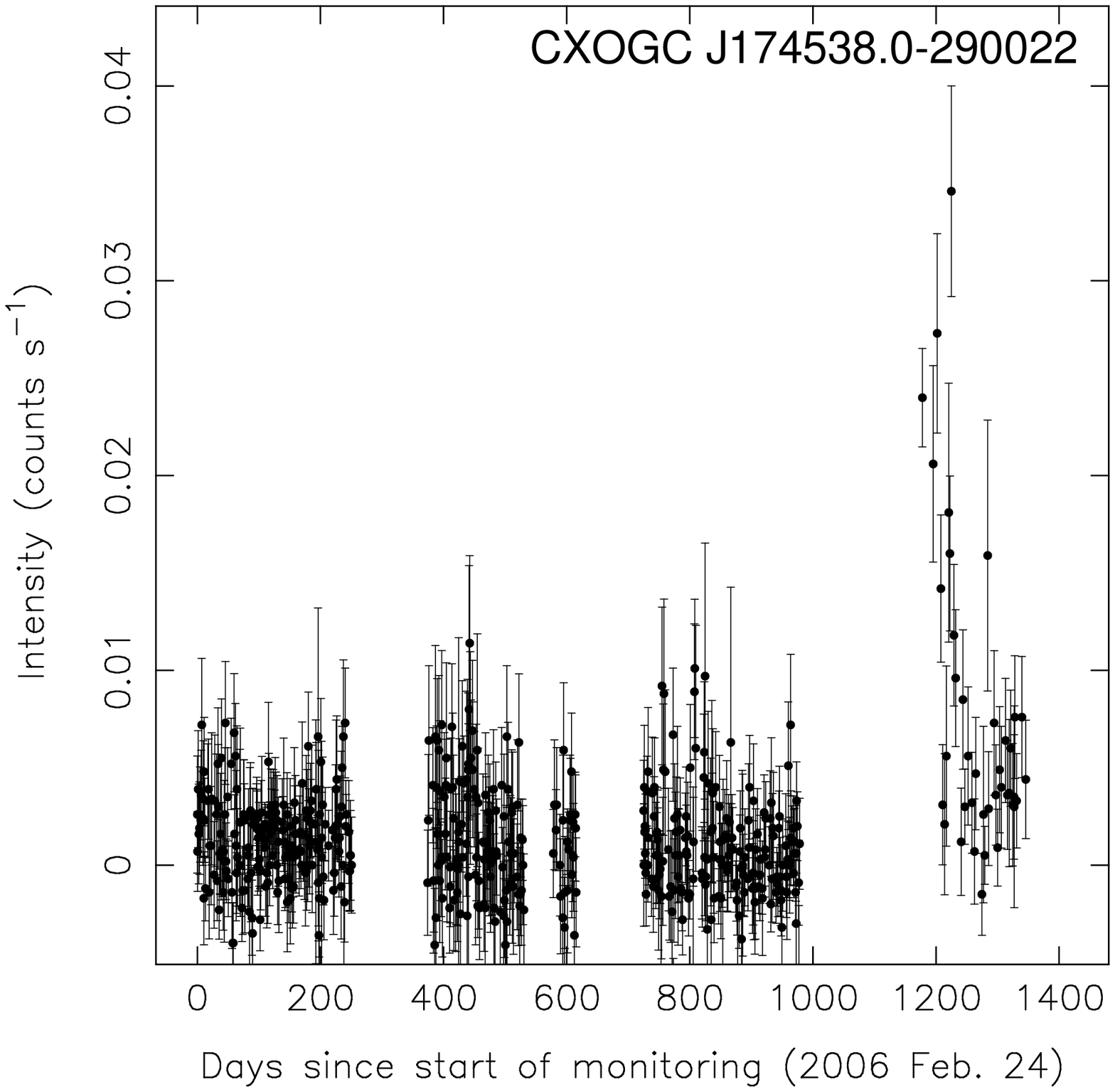}\hspace{2.5cm}
    \end{center}
\caption[]{Background corrected 0.3--10 keV \swift/XRT lightcurves of the five transients that were active in 2008--2009 (PC mode data only). Displayed is their long-term behavior from the start of the monitoring campaign of the GC on 2006 February 24. Days 0--616 cover the years 2006 and 2007 \citep[discussed in][]{degenaar09_gc}, whereas days 725--1346 represent the new data assembled in 2008--2009. The lightcurve of \grsbron\ shows a magnified image of the short and weak outburst that occurred in 2006.}
 \label{fig:lc_long}
\end{figure*}

We constructed X-ray lightcurves for the five active transients using all PC mode observations in which a source was in field of view (FOV). A circular source region with a 5 pixel radius was employed for \brontwee, \xmmbron\ and \bronacht, while a 10 pixel radius was used for the brighter \ascabron\ and \grsbron. Corresponding background events were averaged over a set of three nearby areas having the same shape and size as the source region. Particularly for sources within close proximity of \sgra, which are embedded in diffuse emission (see Fig.~\ref{fig:ds9}), the background regions were carefully chosen to account for the enhanced background level. 
For \ascabron, \grsbron\ and \bronacht\ we could use the full 2008--2009 data set to create the X-ray lightcurve. However, for both \brontwee\ (which is only visible when \ascabron\ is quiescent) and \xmmbron\ (which has a large offset from \sgra\ and was therefore not always in FOV), the lightcurve was constructed from a sub-set of the observations. We combined the 2008--2009 data with that of 2006--2007 \citep[discussed in][]{degenaar09_gc}, to obtain lightcurves with a four-year long baseline for the five active transients. These are displayed in Fig.~\ref{fig:lc_long}. 

The PC data of \ascabron\ and \grsbron\ are subject to pile-up. This effect becomes an issue for PC mode count rates above $\sim 0.5~\cnts$ and causes multiple photons to be registered as single events, thus underestimating the true count rate. The lightcurves of \ascabron\ and \grsbron\ are not corrected for this. To estimate to which extend the count rates are affected, we subtracted source events from both circular and annular regions for the piled-up data of both sources. By comparing the fluxes deduced from spectral fitting, we find that pile-up causes the PC mode count rates to be underestimated by a factor of $\sim1.2$ at $\sim0.7~\cnts$, $\sim1.5$ at $\sim1.0~\cnts$ (about the peak count rate of \ascabron) and $\sim3$ at $\sim1.8~\cnts$ (the peak count rate observed for \grsbron).

We combined all the pointings in which a source was active to create a summed outburst spectrum (displayed in Fig.~\ref{fig:spec}). For \ascabron\ and \grsbron, we attempted to circumvent the effect of pile-up on the spectral shape by using an annular extraction region for the PC data, following the \swift\ pile-up analysis thread.\footnote{http://www.swift.ac.uk/pileup.shtml.} As such, we used an annulus with an inner (outer) radius of 6 (15) pixels to extract source events from \ascabron, and an inner (outer) radius of 7 (30) pixels for \grsbron. We compared the pile-up corrected PC spectra of \grsbron\ with spectra obtained from quasi-simultaneous WT mode data, which are not subject to pile-up. For the WT data, we used a $40\times40$ pixels rectangular extraction region for source events, and a box of similar dimensions as a background reference. From spectral fitting, we obtained fluxes for the different modes that differ by $\sim10\%$, and spectral parameters that are consistent with one another within the errors. This suggests that the above described pile-up correction for the PC mode works satisfactory.

Using the software tool \textsc{xrtexpomap}, we generated exposure maps for each observation, which carry information about the bad columns and hence the effective area of the CCD. These were subsequently used to create ancillary response files (arf) for all spectra with the task \textsc{xrtmkarf}. These account for different extraction regions, vignetting and corrections for the point spread function. The latest response matrix files (rmf, v. 11) were taken from the CALDB database. Using \textsc{grppha}, the spectra were grouped to contain at least 20 photons per bin. However, for \xmmbron\ and \brontwee\ we use bins with a minimum of 10 photons, because of the low number of counts collected for these two sources.

We fitted the average outburst spectra using \textsc{Xspec} \citep[v. 12.5;][]{xspec} to an absorbed powerlaw model and deduce the absorbed and unabsorbed fluxes in the 2--10 keV energy range. For the neutral hydrogen absorption, we use the \textsc{phabs} model using the default \textsc{Xspec} abundances and cross-sections. When a source displayed multiple outbursts, we fitted these simultaneously with the hydrogen column density tied. We include the 2006 and 2007 data \citep[discussed in][]{degenaar09_gc} in these fits. To calculate the unabsorbed {\it peak} flux of the outbursts of \ascabron, \xmmbron\ and \grsbron, we extracted a single spectrum from the observation with the highest count rate. For the former two we use PC observations, applying pile-up corrections as described above when necessary, while for the latter source WT data was available. In case of \brontwee\ and \bronacht, the source count rates were too low to extract a spectrum from a single observation. Therefore, we determined a count rate to flux conversion factor for these two sources by comparing the average outburst flux with the average net count rate. We then used this to estimate the unabsorbed peak flux from the observed peak count rate.

All five active X-ray transients are heavily absorbed ($N_{\mathrm{H}}\gtrsim 7 \times 10^{22}~\mathrm{cm}^{-2}$; see Table~\ref{tab:spectra}), consistent with values obtained for sources that lie close to \sgra. We therefore assumed source distances of 8~kpc when calculating 2--10 keV luminosities from the unabsorbed fluxes. However, for \grsbron\ we use a distance of 7.2~kpc, since this is the upper limit that has recently been inferred from the analysis of thermonuclear X-ray bursts \citep{trap09}. The X-ray spectra are displayed in Fig.~\ref{fig:spec} and the results of our spectral analysis are presented in Table~\ref{tab:spectra}. The 2006 and 2007 data \citep[reported by][]{degenaar09_gc}, are refitted in this work. 

For each observed outburst we calculate the fluency by multiplying the average unabsorbed 2--10 keV outburst flux with the duration of the outburst. These results are presented in Table~\ref{tab:mdot}. Assuming that the transients are accreting systems, we additionally calculate the mass-accretion rates during outburst. Furthermore, we use the estimated duty cycles to obtain an order of magnitude approximation for their long-term mass-accretion rates (see Section~\ref{subsec:mdot} for further details). These results are also included in Table~\ref{tab:mdot}.


\section{X-ray lightcurves and spectra}\label{sec:results}

\subsection{\ascabron}\label{subsec:ascabron}
The X-ray transient \ascabron\ was detected with the \textit{ASCA} satellite in 1993 October and in 1994 October, displaying 3--10 keV luminosities of $\sim 2 \times 10^{35}$ and $\sim9 \times 10^{35}~\lum$, respectively \citep[][]{maeda1996}. The detection of thermonuclear X-ray bursts identified the source as a neutron star LMXB and the system displays eclipses with a period of 8.4 h, which represents the binary orbital period \citep[][]{maeda1996}. The likely quiescent counterpart, \ascabroncxo, has a 2--10 keV luminosity of several times $10^{32}~\lum$ \citep[][]{degenaar09_gc}.

Since 1994, \ascabron\ was never reported in outburst again, despite regular monitoring of the GC, e.g., with the \chan\ satellite between 1999 and 2004 \citep[][]{muno03, muno04_apj613, muno05_apj622}. However, the source was found active in 2006 February, when the \swift/XRT monitoring observations of the GC kicked off \citep{kennea06_atel753}.\footnote{The transient was first denoted as Swift J174535.5--290135, but the detection of 8.4 hr eclipses in \xmm\ observations definitely linked Swift J174535.5--290135 to \ascabron\ \citep{porquet07}.} \ascabron\ remained active for 16 weeks at an average 2--10 keV luminosity of $\sim4 \times 10^{35}~\lum$ \citep[][see also Table~\ref{tab:spectra}]{degenaar09_gc}. Subsequently, the source resided in quiescence for at least four months (2006 July--October), but was again detected in outburst by \inte\ and \swift\ on 2007 February 15--17 \citep{kuulkers07_atel1005,wijnands07_atel1006}. The activity continued throughout the 2007 monitoring campaign, which ended on 2007 November 2. The average outburst luminosity was $\sim1.5 \times 10^{36}~\lum$ (2--10 keV), i.e., $\sim4$ times higher than the level observed in 2006 \citep{porquet07,degenaar09_gc}. 

When the \swift/GC monitoring observations resumed on 2008 February 19, \ascabron\ was detected at a similar intensity as measured in 2007 November (see Fig.~\ref{fig:lc_long}). This makes it likely that the outburst continued during the time that \swift\ could not observe the GC due to Sun-angle constraints. In 2008, the source flux was observed to decrease gradually (see Fig.~\ref{fig:lc_long}). In late-August, the decay accelerated and within two weeks the source luminosity dropped from $\sim10^{36}~\lum$ (2--10 keV), down to the background level on 2008 September 2. \ascabron\ was not detected for the remainder of the \swift/XRT observations in 2008 and the system had thus returned to quiescence following an accretion outburst that lasted $>1.5$ years ($>80$~weeks). In 2009, no activity was detected, indicating that the source remained in quiescence. However, in 2010 June the source is again detected in outburst by \swift/XRT at a 2--10 keV luminosity of a few times $10^{35}~\lum$ \citep[][]{degenaar2010_atel_asca}.

The different outburst spectra of \ascabron\ are displayed in Fig.~\ref{fig:spec}. 
Both the 2006 and 2007--2008 outburst have a soft X-ray spectrum with powerlaw indices of $\Gamma=2.4\pm0.1$ and $2.7\pm0.1$, respectively (see Table~\ref{tab:spectra}). The 2006 outburst appears to have a harder X-ray spectrum than the brighter 2007--2008 outburst (we obtain similar results when $N_{\mathrm{H}}$ is left as a free parameter and not fixed between the outbursts). The spectra indicate that the system is heavily absorbed with a best fit hydrogen column density of $N_{\mathrm{H}}=(23.8\pm0.5)\times10^{22}~\mathrm{cm}^{-2}$. The average 2--10 keV unabsorbed flux during the 2007--2008 outburst was $\sim1\times10^{-10}~\flux$. For an outburst duration of $>80$~weeks, this implies a fluency of $\gtrsim7\times10^{-3}~\fluence$, which is a factor of $\sim 10$ higher than that of the 2006 outburst (see Table~\ref{tab:mdot}).


\subsection{\brontwee}\label{subsec:brontwee}
The X-ray source \brontwee\ is located only $\sim 14''$ away from \ascabron\ (see Fig.~\ref{fig:ds9}). This transient was discovered in 2001 during a monitoring campaign of the GC with \chan\ \citep[][]{muno04_apj613}. Since then, the source has been detected in outburst multiple times with \chan, \xmm\ and \swift, displaying typical luminosities of $\sim10^{33-34}~\lum$ \citep[][]{muno05_apj622,wijnands05_atel638,wijnands06_atel892,deeg08_atel_gc,degenaar09_gc}. In quiescence, the source has not been detected, yielding an upper limit on luminosity of $<9 \times 10^{30}~\lum$ \citep[2--8 keV;][]{muno05_apj622}.

\swift/XRT cannot spatially resolve \brontwee\ and \ascabron~when the latter, which is the brightest of the two, is active. We can therefore only deduce information on the activity of \brontwee\ from \swift\ data obtained in epochs that \ascabron\ is quiescent, which is 2006 July--November and 2008 September onwards (see Section~\ref{subsec:ascabron}). In 2006, the \swift/XRT observations captured an outburst from \brontwee\ that had a duration of $>12$~weeks \citep[][]{degenaar09_gc}. 

Renewed activity of \brontwee\ was revealed by \chan\ observations obtained on 2008 May 10--11, when the source displayed a 2--10 keV luminosity of a few times $10^{33}~\lum$ \citep[][]{degenaar08_atel_gc_chan}. After \ascabron\ had returned to quiescence in 2008 September, \brontwee\ was continuously detected until the monitoring observations ended on 2008 October 30. This outburst thus lasted for $>8$~weeks. If the activity observed by \swift/XRT was part of the same outburst that was detected by \chan\ in 2008 May, the outburst duration increases to $>24$~weeks. Alternatively, if the source returned to quiescence in between, the dormant phase must have lasted $<18$~weeks. The source was not found active during the 2009 observations (see Fig.~\ref{fig:lc_long}).

\begin{figure*}
 \begin{center}
      \includegraphics[width=5.4cm]{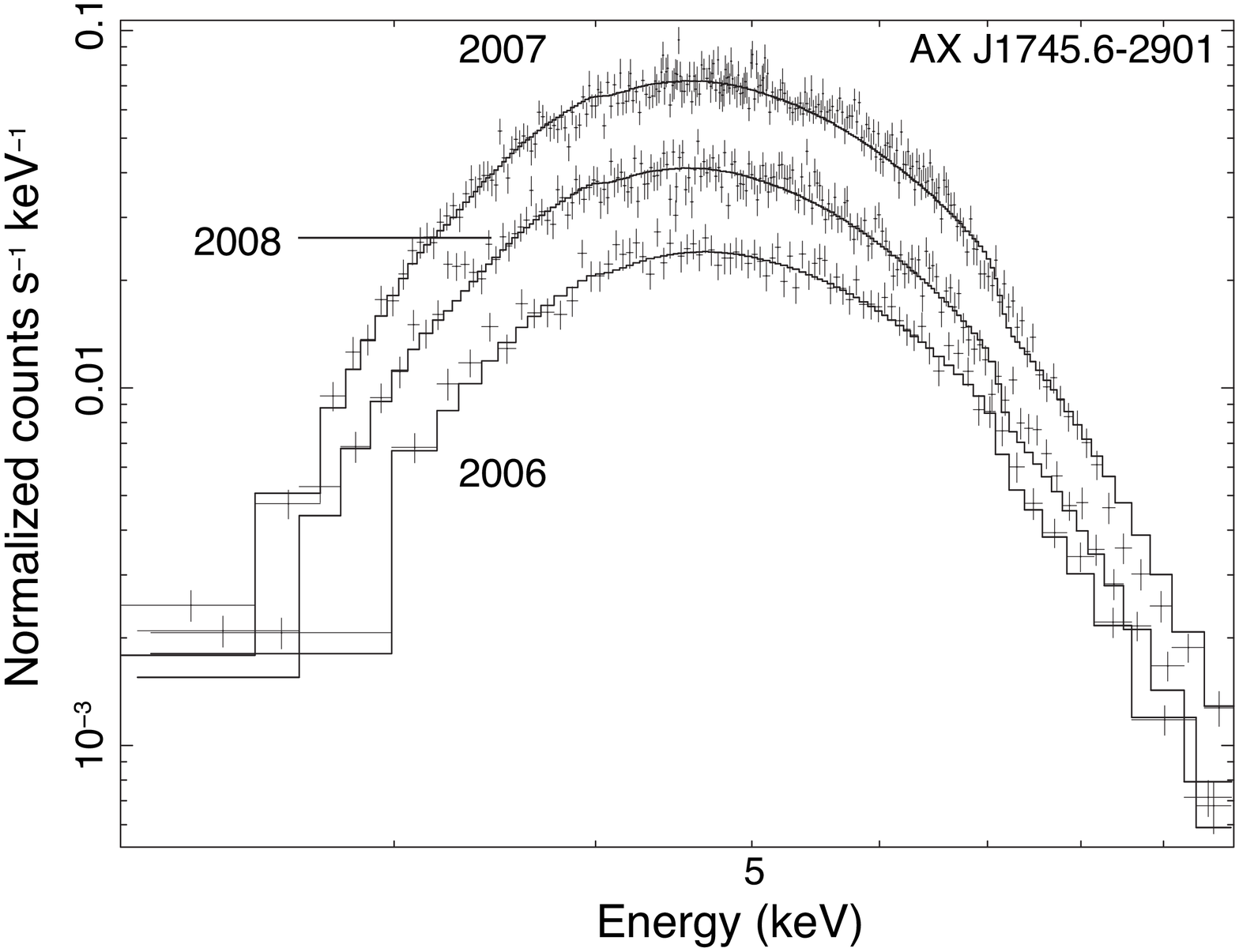}
          \hspace{0.5cm}\vspace{0.3cm}
    \includegraphics[width=5.4cm]{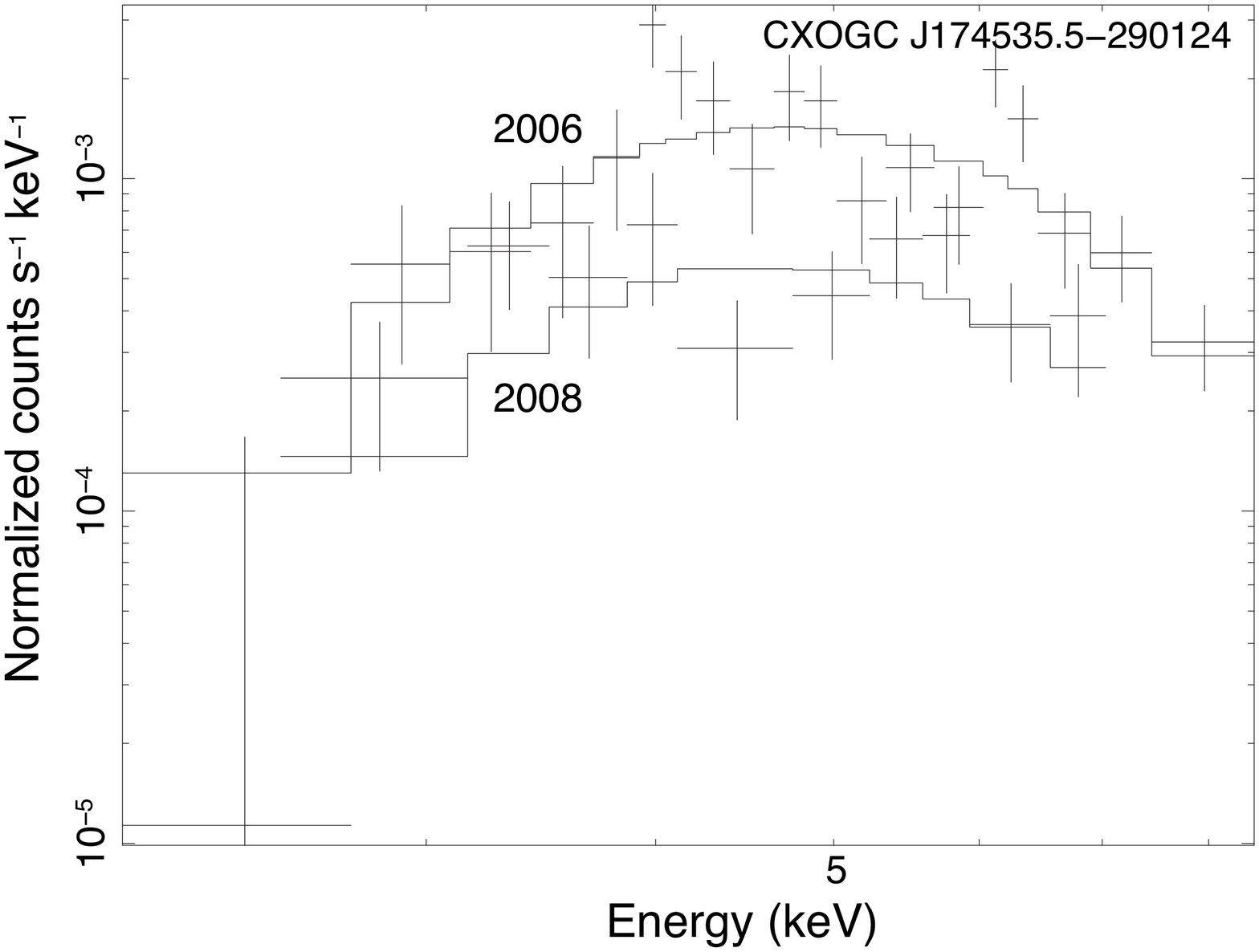}
    \hspace{0.5cm}     
    \includegraphics[width=5.4cm]{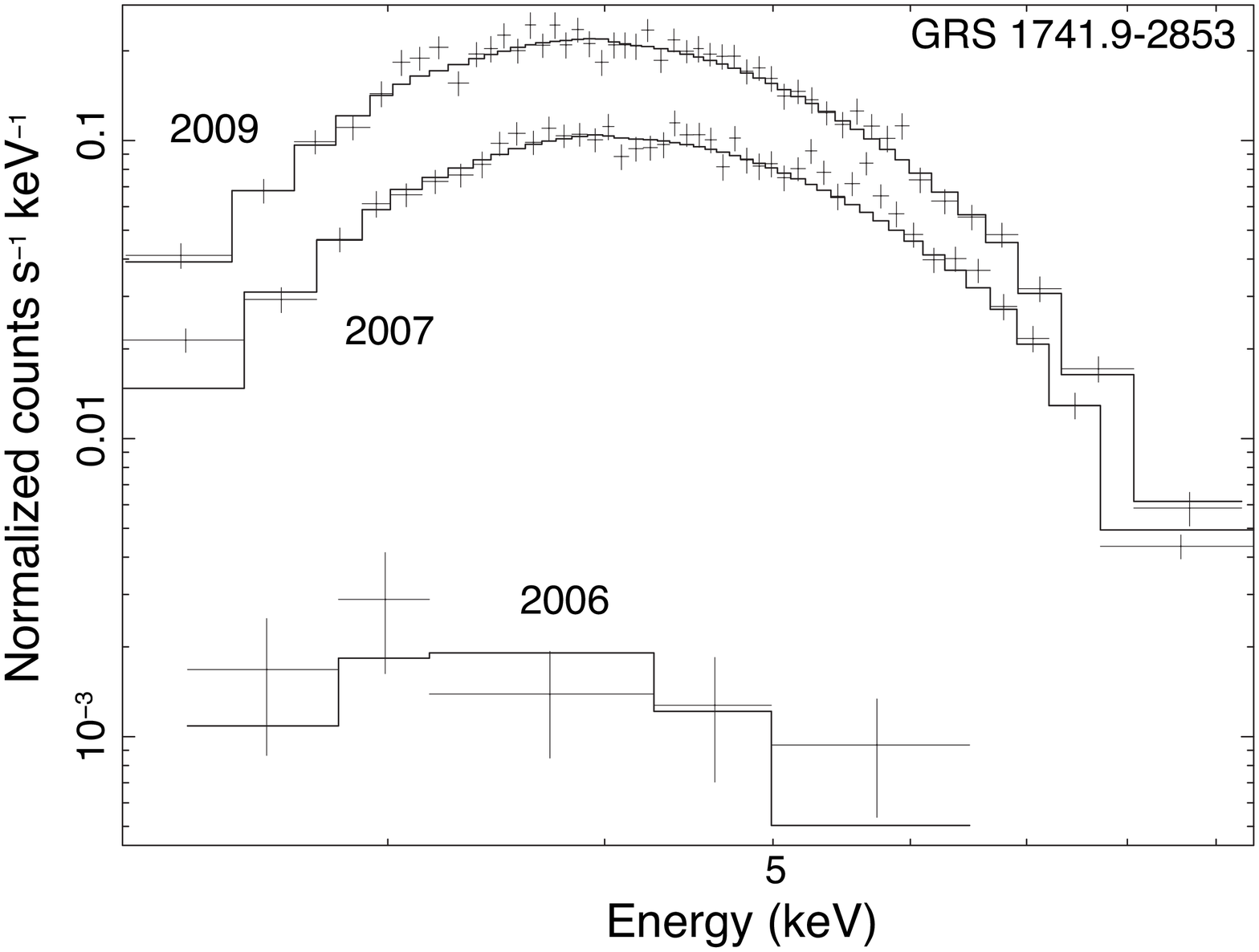}
         \hspace{0.5cm}     
    \includegraphics[width=5.4cm]{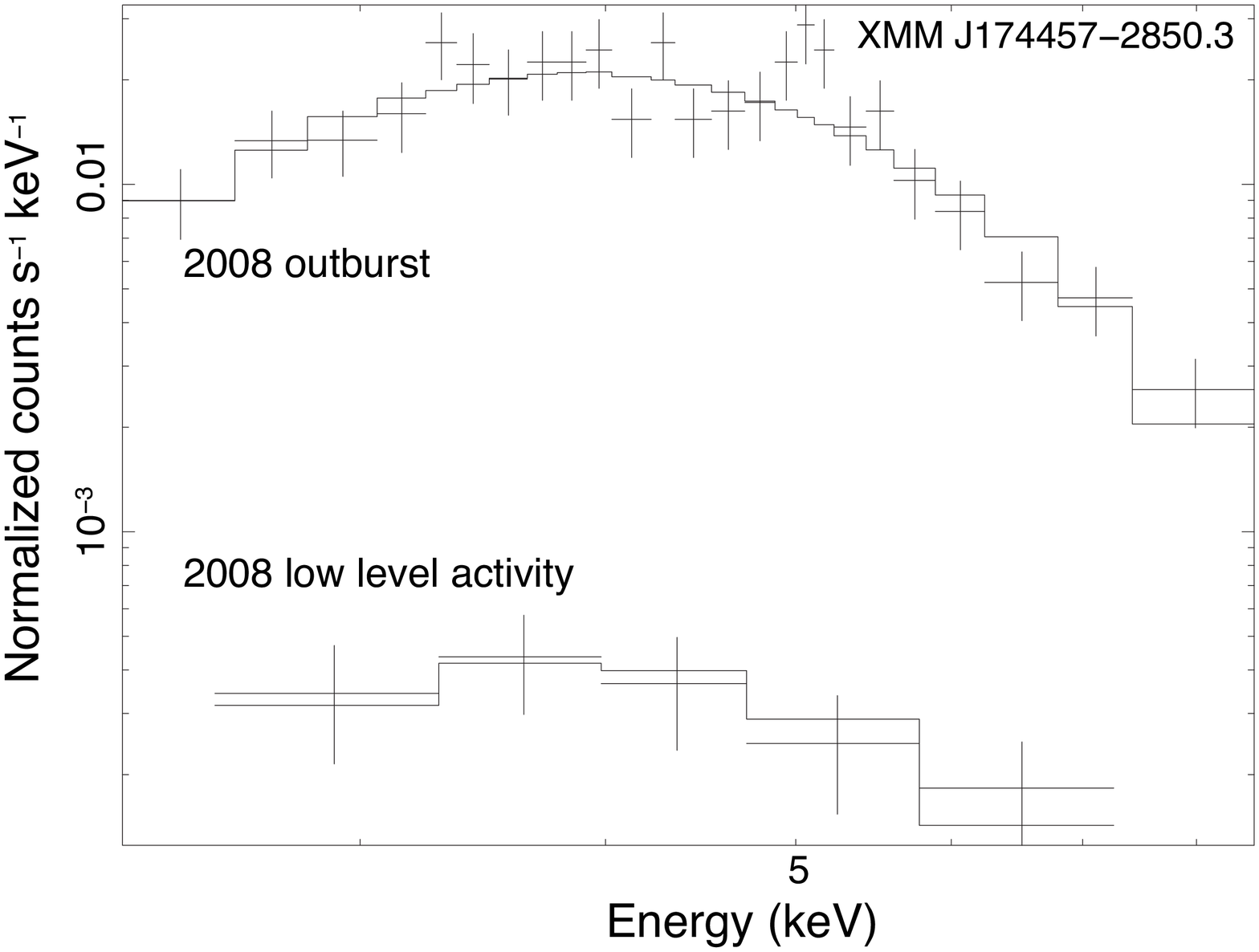}
    \hspace{0.5cm}
           \includegraphics[width=5.4cm]{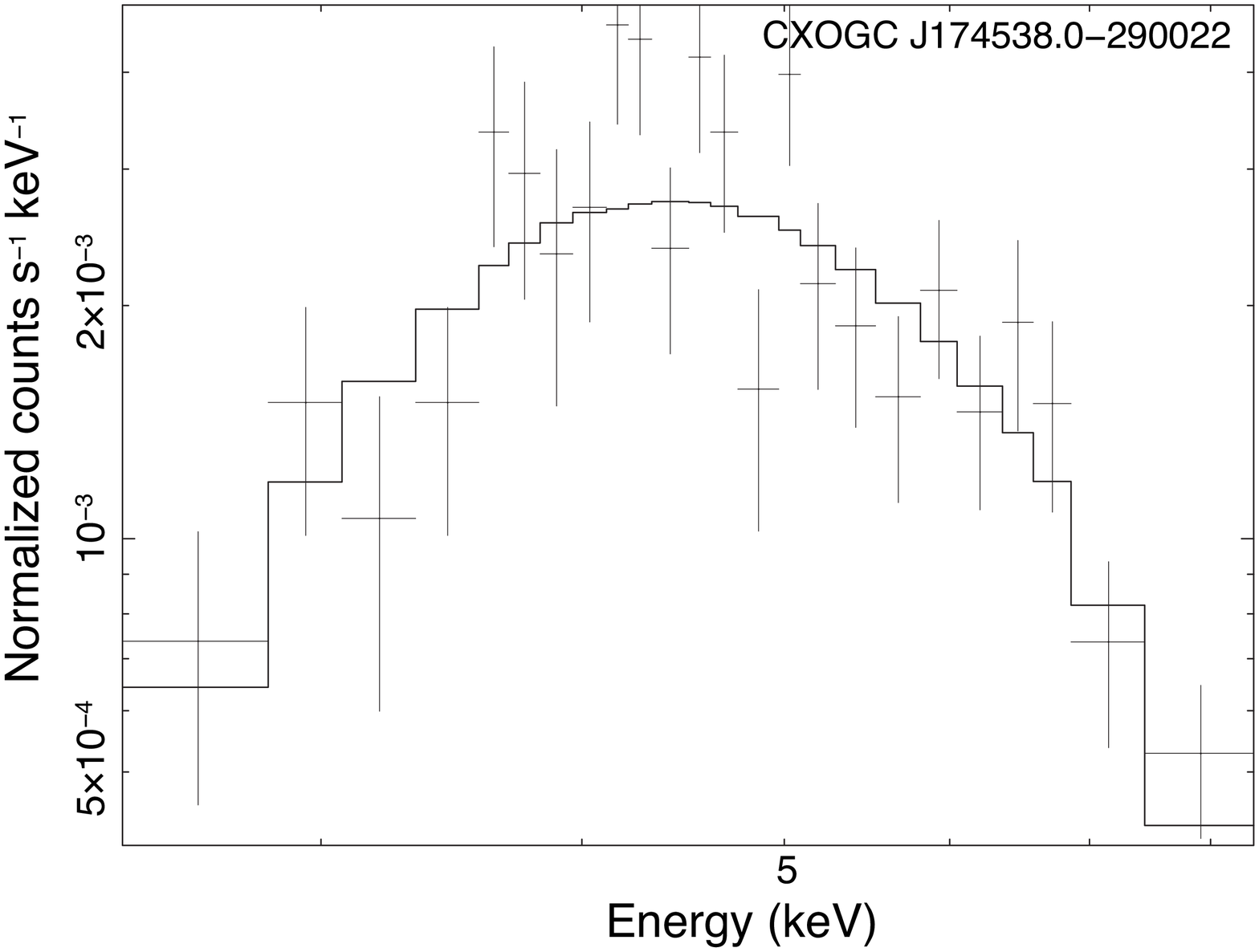}
    \end{center}
\caption[]{Background corrected \swift/XRT average outburst spectra. For \ascabron, \brontwee\ and \grsbron, spectra of the different outbursts captured during the 2006--2009 \swift/XRT monitoring are plotted together. The plot of \xmmbron\ shows the average outburst spectrum of the source in 2008, as well as the spectrum of the low-level activity observed in the months following this outburst.}
 \label{fig:spec}
\end{figure*}

The average 2--10 keV unabsorbed flux observed with \swift/XRT in 2008 was $\sim 1 \times 10^{-12}~\flux$, a factor $\sim1.5$ lower than observed in 2006 (see Table~\ref{tab:spectra}). For an outburst duration of $>8$~weeks, we can constrain the fluency of the 2008 outburst to be $>7\times10^{-6}~\fluence$ (2--10 keV). This increases to $>2\times10^{-5}~\fluence$ if the outburst endured for more than 24~weeks. The spectrum of \brontwee\ is heavily absorbed ($N_{\mathrm{H}}=(12.0\pm6.9) \times10^{22}~\mathrm{cm}^{-2}$) and for both outbursts we obtain a rather hard spectral index of $\Gamma \sim1$, although the uncertainties on this parameter are very large (see Table~\ref{tab:spectra}). The X-ray spectra of the 2006 and 2008 outbursts are shown in Fig.~\ref{fig:spec}.

\subsection{\grsbron}\label{subsec:grsbron}
\grsbron\ was discovered in 1990 March--April by the \textit{Granat} satellite \citep[][]{sunyaev1990}. Since then, the system has been detected in outburst on multiple occasions with 2--10 keV luminosities of $\sim10^{36}~\lum$ \citep[e.g.,][]{sakano02,muno03_grs,wijnands06,trap09}. In quiescence the source displays a luminosity of $\sim10^{32}~\lum$ \citep[][]{muno03_grs}. The detection of thermonuclear X-ray bursts by \textit{BeppoSAX} established that this system is an LMXB harboring a neutron star \citep[][]{cocchi99}.

Between 2006 September 14--20, the \swift/XRT monitoring observations detected a short ($\sim$1 week) outburst from \grsbron, which reached a 2--10 keV peak luminosity of only $\sim 9 \times 10^{34}~\lum$. This is significantly lower than other outbursts observed from this source ($L_X\sim 10^{36-37}~\lum$). In 2007 February--April, the source experienced a longer ($>$13 weeks) and brighter (2--10 keV peak luminosity of $\sim 2 \times 10^{36}~\lum$) accretion outburst, which was captured by several satellites \citep[][]{kuulkers07_atel1008,porquet07,muno07_atel1013,degenaar09_gc}. 

\grsbron\ remained dormant throughout the 2008 \swift\ monitoring campaign (see Fig.~\ref{fig:lc_long}), but experienced another accretion outburst in 2009 October, which was registered by \inte\ and \swift\ \citep[][]{chenevez09,kennea09}. While \grsbron\ was not detected during \swift/XRT observations performed on 2009 September 23, it was found active on 2009 September 29 and the flux started to rise in the following days \citep[][see also Fig.~\ref{fig:lc_long}]{kennea09}. The source intensity had decayed to the background level during the last observation of the campaign, performed on 2009 November 1. This suggests that the 2009 outburst had a duration of 4--5~weeks. The outburst reached a peak luminosity of $\sim1 \times10^{37}~\lum$, while the average outburst value was $\sim2 \times10^{36}~\lum$ (2--10 keV; see Table~\ref{tab:spectra}). To our knowledge, this is the highest peak luminosity ever reported for \grsbron. 

By fitting the data of the three outbursts simultaneously, we obtain a hydrogen column density of $N_{\mathrm{H}}=(14.0\pm0.7)\times10^{22}~\mathrm{cm}^{-2}$. The X-ray spectra, displayed in Fig.~\ref{fig:spec}, are soft with powerlaw indices of $\Gamma=5.0\pm2.5$, $2.6\pm0.1$ and $3.0\pm0.2$ for the 2006, 2007 and 2009 activity, respectively. We note that our values for both $N_{\mathrm{H}}$ and $\Gamma$ are a bit higher than obtained from \chan\ and \xmm\ observations, albeit largely consistent within the errors \citep[][]{muno03_grs,wijnands06,trap09}. The uncertainty on the spectral index of the short and weak 2006 outburst is very large, but comparing the 2007 and 2009 outburst data suggests that the X-ray spectrum is softer for the brightest of the two outbursts. Using a duration of 4--5 weeks and an average unabsorbed flux of $\sim2.8\times 10^{-10}~\flux$ (see Table~\ref{tab:spectra}), we can estimate that the 2009 outburst of \grsbron\ had a 2--10 keV fluency of $\sim 8 \times 10^{-4}~\mathrm{erg~cm}^{-2}$. Despite the different outburst duration and peak luminosity, this is comparable to the 2007 outburst observed by \swift/XRT (see Table~\ref{tab:mdot}).

\begin{table*}
\label{tab:spec}
\begin{threeparttable}[t]
\begin{center}
\caption[]{{Results from fitting the X-ray spectral data.}}
\begin{tabular}{l l l l l l l l l l l}
\hline
\hline
Source name & Year &$N_{\mathrm{H}}$ & $\Gamma$ & red. $\chi^{2}$ & $F_{\mathrm{X, abs}}$ & $F_{\mathrm{X, unabs}}$ & $F_{\mathrm{X, peak}}$ & $L_{\mathrm{X}}$ & $L_{\mathrm{X, peak}}$ \\
 & & ($10^{22}~\mathrm{cm}^{-2}$) &  & (d.o.f.) &  &  &  &  & \\
\hline 
\ascabron & & $23.8\pm0.5$ &  & 1.10 (1427) &  &  &  &  \\
& 2008 & & $2.7\pm 0.1$ &  & $22.5\pm0.2$ & $93.2 \pm 5.5$ & 710 & 71 & 550 \\
& 2007 &  & $2.7 \pm 0.1$ &  & $ 45.2\pm0.3$ & $187.7\pm7.5$ & $800$ & $145$ & $610$  \\
& 2006 &  & $2.4 \pm 0.1$ &  & $14.8\pm0.2$ & $53.3\pm2.2$ & $120$ & $41$ & $92$ \\
\brontwee & & $12.0\pm6.9$ &  & 1.23 (25) &  &  &  &   \\
& 2008 &  & $1.0^{+2.2}_{-1.0}$ &  & $0.9\pm0.3$ & $1.4\pm0.6$ & 2.6 &  1.1 & 2.0  \\
& 2006 &  & $0.8^{+1.1}_{-0.8}$ &  & $ 1.3\pm0.2$ & $2.1\pm0.6$ & $4.0$ & $1.6$ & $3.0$ \\
\grsbron & & $14.0\pm0.7$ &  & 0.99 (416) &  &  &  &   \\
& 2009 &  & $3.0\pm0.2$ &  & $87.2\pm1.5$ & $283.6\pm2.2$ & $2200$ & $176$ & $1300$ \\
& 2007 &  & $2.6\pm0.1$ &  & $61.7\pm1.0$ & $174.9\pm1.2$ & 260 & 109 & 150  \\
& 2006 &  & $5.0\pm2.5$ &  & $0.6\pm0.3$ & $5.0\pm3.7$ & 12 & 3.1 & 7.0  \\
\xmmbron & & $7.5\pm2.9$ &  & 0.83 (29) &  &  &  &  \\
& 2009 &  & $2.3\pm1.1$ &  & $11.5\pm1.8$ & $21.3\pm6.4$ & $21.8$ & $16$ & $17$  \\
& 2008 &  & $1.6\pm0.6$ &  & $20.7\pm2.0$ & $32.4\pm5.4$ & 250 & 25 & 190  \\
& 2008-low &  & $1.8\pm1.4$ &  & $0.5\pm0.2$ & $0.8\pm0.2$ & 3.6 & 0.6 &  2.8  \\
& 2007 & $7.5$ fix& $1.8$ fix&  & $0.4$ & $0.6$ & $1.5$ & $0.4$ & $1.1$  \\
\bronacht & 2009 & $12.8\pm5.9$ & $1.4\pm0.9$ & 1.14 (21) & $2.7\pm0.5$ & $5.0\pm1.4$ & $21.8$ & $3.8$ & $17$ \\
\hline
\end{tabular}
\label{tab:spectra}
\begin{tablenotes}
\item[]Note. -- For sources that displayed multiple outbursts, we fitted the different outburst spectra simultaneously with the hydrogen column density tied. The 2006 and 2007 data \citep[discussed by][]{degenaar09_gc} are re-fitted in this work. Fluxes and luminosities are for the 2--10 keV energy band and given in units of $10^{-12}~\mathrm{erg~cm}^{-2}~\mathrm{s}^{-1}$ and $10^{34}~\mathrm{erg~s}^{-1}$, respectively. $F_{\mathrm{X, abs}}$ and $F_{\mathrm{X, unabs}}$ represent the mean absorbed and unabsorbed outburst fluxes, while $F_{\mathrm{X, peak}}$ is the unabsorbed peak flux. $L_{\mathrm{X}}$ and $L_{\mathrm{X, peak}}$ are the average and peak outburst luminosity, respectively. These are calculated from the unabsorbed fluxes by adopting a distance of 7.2~kpc for \grsbron\ and 8 kpc for all other sources. Fluxes for the 2007 activity of \xmmbron\ were deduced using \textsc{pimms}, for fixed values of $N_{\mathrm{H}}$ and $\Gamma$.
\end{tablenotes}
\end{center}
\end{threeparttable}
\end{table*}


\subsection{\xmmbron}\label{subsec:xmm}
\xmmbron\ is a transient X-ray source that was first detected in outburst with \xmm\ in 2001 September, when it displayed a 2--10 keV luminosity of $\sim 5\times10^{34}~\lum$   \citep[][]{sakano05}. Since its initial discovery, \xmmbron\ has been active repeatedly, displaying 2--10 keV X-ray luminosities in a broad range of a few times $\sim 10^{33}~\lum$, up to $\sim 10^{36}~\lum$ \citep[][]{wijnands06,muno07_atel1013}. 

As mentioned earlier, \xmmbron\ is only in FOV in a sub-set of the \swift/XRT monitoring data, due to its relatively large offset from \sgra\ ($\sim 13.7'$). The source was never in FOV during the 2006 observations. In 2007, the source field was covered a few times between July and November, and \xmmbron\ was detected at 2--10 keV luminosities of $\sim 10^{33-34}~\lum$ \citep[][]{degenaar09_gc}.
When the source first came into view in 2008, on June 28, it displayed a 2--10 keV X-ray luminosity of $\sim 1 \times 10^{36}~\lum$. The source intensity decreased over the course of a few days, down to a level of a few times $10^{33}~\lum$ around 2008 July 7 (see Fig.~\ref{fig:lc_long}). 

Following this decay, the source remained to be detected by \swift/XRT all through the end of the monitoring observations on 2008 October 30. During this episode, \xmmbron\ displayed a 2--10 keV luminosity of a few times $\sim10^{33}~\lum$, which is a factor $>10$ above its quiescent level of $\sim10^{32}~\lum$ \citep[][]{sakano05}. We extracted separate spectra of the bright outburst (2008 June 28--July 7), as well as the low-level activity that followed (2008 July 8--October 30). Both spectra are shown in Fig.~\ref{fig:spec} and the spectral parameters and fluxes are listed in Table~\ref{tab:spectra}. We note that the source is not detected in our \chan\ monitoring observations of the GC performed on 2008 May 10 (Degenaar et al. in preparation), which implies that the 2--10 keV luminosity of \xmmbron\ was lower than a few times $10^{33}~\lum$ at that time. The bright active state (2--10 keV luminosity of $\sim10^{35-36}~\lum$) detected in 2008 late-June thus lasted $<49$~days ($<7$~weeks). 

In 2009, the source was detected during a single \swift\ pointing performed on September 29, at a luminosity of $\sim2 \times 10^{35}~\lum$ (2--10 keV). The spectrum of this observation largely overlays the average spectrum of the 2008 outburst and is therefore not plotted in Fig.~\ref{fig:spec}. \xmmbron\ is not active in the preceding, nor in the subsequent observation, carried out on September 23 and October 2, respectively. This implies that the activity lasted less than 9 days.

For the different outbursts captured by \swift\ between 2006--2009, we obtain spectral parameters that are comparable to the outburst values reported by \citet{sakano05} using \xmm\ and \chan\ data obtained in 2001. The source is heavily absorbed ($N_{\mathrm{H}}=(7.5\pm2.9)\times10^{22}~\mathrm{cm}^{-2}$) and the powerlaw index adapts values of $\Gamma \sim1.5-2.5$, with large uncertainties due to the low statistics (see Table~\ref{tab:spectra}). There is no obvious correlation between the spectral index and the source flux. The 2--10 keV fluency of the different outbursts of \xmmbron\ varies between $\sim(0.4-10)\times10^{-5}~\fluence$ (see Table~\ref{tab:mdot}), with an average value of $\sim2.5\times10^{-5}~\fluence$.


\subsection{\bronacht}\label{subsec:bronacht}
The X-ray source \bronacht\ was discovered during \chan\ monitoring observations of the GC \citep{muno03}. Between 1999 and 2004, \chan\ detected the source at a minimum and maximum luminosity of $\sim 1 \times 10^{33}$ and  $\sim 3 \times 10^{34}~\lum$, respectively \citep[2--8 keV;][]{muno05_apj622}. The source was not detected during the \swift/XRT monitoring observations carried out in 2006, 2007 and 2008 (see Fig.~\ref{fig:lc_long}). 

\bronacht\ was reported active as seen during \xmm\ observations obtained between 2009 April 1--5,  displaying a 2--10 keV luminosity of $\sim 2\times10^{34}~\lum$ \citep{ponti09}. A $\sim 4.6$~ks \swift/XRT ToO pointing performed on 2009 May 17, about 6 weeks after the \xmm\ detection, found the source still in outburst. The \swift\ monitoring observations of the GC resumed on 2009 June 4, and the source is clearly detected by visual inspection until 2009 mid-July. The average luminosity during this episode is $\sim9 \times 10^{34}~\lum$, peaking at $\sim2 \times 10^{35}~\lum$ (2--10 keV). 

Although no clear source activity is apparent in individual exposures obtained after 2009 mid-July, summing the data from this period till the end of the monitoring observations (2009 November 1), does result in a weak detection of the source. The average luminosity during this episode is $\sim1 \times 10^{34}~\lum$. Fig.~\ref{fig:spec} displays the average \swift/XRT spectrum of the entire active period, i.e., from 2009 May--November. The spectral parameters and fluxes are listed in Table~\ref{tab:spectra}. 

The outburst captured by \swift\ had a duration of $>9$~weeks. It is likely that the source was continuously active between the \xmm\ detection and the \swift/XRT observations, which would imply that the outburst had a duration of $>30$~weeks. We note that the source is not detected during the 2010 monitoring observations, which commenced on 2010 April 4. This implies that the outburst of \bronacht\ was shorter than 52 weeks. The spectral parameters deduced from fitting the XRT data ($N_{\mathrm{H}}=(12.8\pm5.9)\times10^{22}~\mathrm{cm}^{-2}$ and $\Gamma=1.4\pm0.9$; see Table~\ref{tab:spectra}) are similar to the values inferred from \xmm\ observations obtained in 2008 early-April \citep{ponti09}. If we assume an outburst duration of 30--52~weeks, the average unabsorbed flux of $\sim5\times10^{-12}~\flux$ implies a 2--10 keV outburst fluency of $\sim(9-20)\times10^{-5}~\fluence$ (see Table~\ref{tab:mdot}).


\section{Discussion}\label{sec:discuss}
In this work, we analyzed the lightcurves and spectra of five different X-ray transients that were found active during \swift/XRT monitoring observations of the GC carried out in 2008 and 2009. The four sources \ascabron, \brontwee, \grsbron\ and \xmmbron\ were also active between 2006 and 2007, while \bronacht\ was detected for the first time with \swift\ in 2009. 
The two brightest transients, \ascabron\ and \grsbron, are both confirmed neutron star LMXBs (based on the detection of thermonuclear X-ray bursts), while the other three are unclassified X-ray sources. 

\ascabron\ was observed to return to quiescence in 2008 September, following an unusually long accretion episode that started before 2007 February and endured for $\gtrsim 1.5$~years. In 2009, the \swift/XRT observations captured the brightest outburst ever reported for \grsbron, which reached up to a 2--10 keV peak luminosity of $1.3 \times 10^{37}~\lum$. Both sources appear to have rather soft X-ray spectra with powerlaw indices that are higher ($\Gamma\sim2.5-3.0$; see Table~\ref{tab:spectra}) than typically found for brighter neutron star LMXBs ($\Gamma \sim2$). Furthermore, the 2008--2009 data set reveals that although \xmmbron\ exhibits outbursts with peak luminosities around $\sim 1 \times 10^{36}~\lum$, it can also spend long episodes at a much lower active level of $\sim 10^{33-34}~\lum$ (2--10 keV). 

\brontwee\ and \bronacht\ both display very low 2--10 keV peak luminosities of $\sim 10^{34-35}~\lum$ and have never been detected at higher levels. \brontwee\ was active in 2008, while a previous outburst was detected with \swift/XRT in 2006 \citep[][]{degenaar09_gc}. This confirms that this system has a relatively high duty cycle.
\bronacht\ displayed an outburst peak luminosity of $\sim2 \times 10^{35}~\lum$ (2--10 keV), which is $\sim2$ orders of magnitude higher than the lowest luminosity detected during \chan\ observations of the GC carried out between 1999 and 2004 \citep[$\sim1 \times 10^{33}~\lum$ in the 2--8 keV band;][]{muno05_apj622}. This unambiguously demonstrates the transient nature of this source. If \bronacht\ is an X-ray binary and its quiescent luminosity is $\sim1 \times 10^{33}~\lum$ \citep[][]{muno05_apj622}, this would favor a neutron star as the compact primary, since black hole systems are typically fainter in their quiescent states unless the orbital period is several days \citep[e.g.,][]{narayan97,menou99,lasota07}. 

In 2006, the \swift/XRT monitoring campaign of the GC detected activity of three other transients, \brondrie, \bronvier\ (likely associated to \bronviercxo) and \bronvijf\ \citep[][]{degenaar09_gc}. The former two both experienced short outbursts ($\sim 2$ weeks) with 2--10 keV peak luminosities of $\sim 2\times10^{35}~\lum$. These two sources are not detected during the 2008--2009 observations, which confirms that these systems have low duty cycles \citep[][see also Table~\ref{tab:mdot}]{degenaar09_gc}. The newly discovered transient \bronvijf\ was active for $\sim5$~weeks in 2006 and reached a peak luminosity of $\sim 7\times10^{34}~\lum$ (2--10 keV). This source has a relatively large offset from \sgra\ ($\sim11'$) and was only in FOV during 39 and 4 pointings in 2008 and 2009, respectively. No activity is detected from the source during these observations. 

\subsection{Peculiar source properties}\label{subsec:properties}
\subsubsection{Lightcurve morphology of \ascabron}\label{subsubsec:ascabron}
As discussed in Section~\ref{subsec:ascabron}, the \swift/XRT observations of the GC exposed two distinct outbursts from \ascabron\ between 2006 and 2009, which are very different in terms of duration and luminosity (see Tables~\ref{tab:spectra} and~\ref{tab:mdot}). Since \ascabron\ is a confirmed neutron star LMXB, the disk instability model is thought to provide the framework to explain the outburst behavior of this source.
 
The average 2--10 keV luminosity during the 2007--2008 outburst was $\sim 1\times 10^{36}~\lum$ (see Table~\ref{tab:spectra}). Assuming that the bolometric luminosity is a factor of $\sim 3$ higher \citep[e.g.,][]{zand07}, this implies a mass-accretion rate of $\langle \dot{M}\rangle _{\mathrm{ob}} \sim 3 \times 10^{-10}~\mdot$ for a canonical neutron star with $M=1.4~\Msun$ and $R=10$~km. For an outburst duration of $1.5$~years, this corresponds to a total accreted disk mass of $\sim5\times10^{-10}~\Msun$. In 2006, the outburst had an average luminosity of $\sim 4\times 10^{35}~\lum$ and a duration of $>16$~weeks, which would translate into a mean mass-accretion rate of $\langle \dot{M}\rangle _{\mathrm{ob}} \sim 1 \times 10^{-10}~\mdot$ and a total accreted mass of $\gtrsim3\times10^{-11}~\Msun$.

Given the fact that \ascabron\ is transient, the mass-transfer rate from the companion star must be lower than the accretion rate onto the compact object during outburst \citep[e.g.,][]{king98}, i.e., $\langle \dot{M}\rangle _{\mathrm{tr}} \lesssim 1 \times 10^{-10}~\mdot$ (as estimated from the 2006 outburst; see Table~\ref{tab:mdot}). If the mass-transfer rate from the companion star does not change considerably over time, it would thus take the system at least 5 years to build up the accretion disk that powered the 2007--2008 outburst from scratch. This is consistent with the fact that no similarly long outbursts from this source have been observed between 1994 and 2006 (see Section~\ref{subsec:ascabron}). The duty cycle of similar 1.5-year long outbursts from this system would thus be $\lesssim23\%$. This is in agreement with observational constraints, which result in an estimated duty cycle of $10-30\%$ \citep[][]{degenaar09_gc}. Given the time required to build up an accretion disk that can account for the 2007--2008 activity, and the observed quiescence interval between the 2006 and 2007--2008 outbursts of only $\sim4-7$~months \citep[][see also Fig.~\ref{fig:lc_long}]{degenaar09_gc}, it seems that a significant residual accretion disk must have remained after the 2006 outburst ended. Shorter outbursts like the one observed in 2006 consume much less disk mass and could recur on a timescale of a only a few months. As mentioned in Section~\ref{subsec:ascabron}, \ascabron\ was again reported in outburst in 2010 June, displaying a similar intensity level as in 2006 \citep[][]{degenaar2010_atel_asca}.

Within the disk instability model, we can understand the observed behavior if in 2006 only part of the accretion disk became ionized, while the 2007--2008 outburst drained a larger part of (or maybe the full) accretion disk \citep[see, e.g.,][]{king98,lasota01}. This picture might also provide an explanation for the fact that the 2006 outburst was fainter than the one observed in 2007--2008, since the mass-accretion rate (and thus the accretion luminosity) is expected to scale with the size of the hot ionized zone of the accretion disk \citep[see][]{king98}. We note that the disk instability model for accreting white dwarfs predicts alternating sequences of outbursts with different duration and brightness, consistent with observations of dwarf novae eruptions \citep[e.g.,][]{cannizzo1993,lasota01}. While driven by the same underlying mechanism, it is thought that in LMXBs the stability properties are strongly influenced by irradiation of the accretion disk \citep[][]{king98,lasota01}. As a consequence, LMXBs are expected to consume a larger part of the accretion disk during outbursts, which are therefore longer and less frequent than observed for dwarf novae \citep[][]{king98,lasota01}.

\subsubsection{Recurrence time of \grsbron}\label{subsubsec:grsbron}
Despite the different duration and average flux, the fluency of the 2009 outburst of \grsbron\ is comparable to the 2007 outburst fluency of $\sim 1 \times 10^{-3}~\mathrm{erg~cm}^{-2}$ (see Table~\ref{tab:mdot}). The two outbursts are separated by an epoch of $\sim2.5$~year. The total mass accreted during the 2007 outburst can be estimated as $\gtrsim7\times10^{-11}~\Msun$ (for $t_{\mathrm{ob}}>13$~weeks and $\langle \dot{M}\rangle_{\mathrm{ob}} \sim 3 \times 10^{-10}~\mdot$). For the 2009 outburst we obtain a comparable value of $\sim4\times10^{-11}~\Msun$ ($t_{\mathrm{ob}}\sim4-5$~weeks and $\langle \dot{M}\rangle_{\mathrm{ob}} \sim 5 \times 10^{-10}~\mdot$). 

In 2005, \grsbron\ also underwent an accretion outburst that endured for several weeks. The rise of this outburst was captured by \inte\ in 2005 early-April \citep{kuulkers07}, while \chan\ observations indicated that the source was fading in 2005 early-July \citep{wijnands06}. This suggests an outburst duration of $\sim13$~weeks. Assuming an average 2--10 keV flux of $\sim1\times10^{-10}~\flux$ \citep[as inferred from \chan\ observations performed in 2005 June;][]{wijnands06}, we can asses that the 2005 outburst had a 2--10 keV fluency of approximately $\sim 8 \times 10^{-4}~\mathrm{erg~cm}^{-2}$. This is comparable to the two large outbursts occurring in 2007 and 2009 (see Table~\ref{tab:mdot}). The time between the 2005 and the 2007 outburst is nearly 2 years. The total mass accreted during the 2005 outburst can be estimated as $\sim5\times10^{-11}~\Msun$ (assuming $t_{\mathrm{ob}}\sim13$~weeks and $\langle \dot{M}\rangle_{\mathrm{ob}} \sim 2 \times 10^{-10}~\mdot$). This is very similar to the values estimated above for the 2007 and 2009 outbursts.

Based on the three outbursts observed for \grsbron\ in the past 5 years (2005, 2007 and 2009, neglecting the weak and short outburst captured by \swift\ in 2006) we can infer that the system has typical outburst duration on the order of $\sim10$~weeks. The detection history in the past decade \citep[see][for an overview]{trap09}, suggests a recurrence time of roughly $\sim 2$~years. This implies a duty cycle of $\sim10\%$ and the average accretion rate during outburst appears to be typically a few times $10^{-10}~\mdot$.

\subsubsection{\xmmbron: a wind-fed system?}\label{subsubsec:xmmbron}
As discussed in Section~\ref{subsec:xmm}, the unclassified transient X-ray source \xmmbron\ has a quiescent level of $L_{X}\sim 10^{32}~\lum$, while the observed maximum luminosity is $\sim 10^{36}~\lum$ (2--10 keV). The 2008 \swift/XRT observations of the GC show that the bright stages of this source might only last for a few days, while \xmmbron\ is often found at levels intermediate between quiescence and full outburst, at a 2--10 keV luminosity of $\sim 10^{33-34}~\lum$. Such behavior is difficult to understand within the framework of accretion disk instabilities in LMXBs. Instead, wind accretion might provide a more natural explanation.

The activity displayed by \xmmbron\ is in some ways reminiscent of the behavior observed from SFXTs, which harbor neutron stars accreting from the stellar wind of a supergiant O/B companion \citep[e.g.,][]{negueruela06}. These systems undergo sporadic X-ray flares lasting only a few hours to days and reaching up to 2--10 keV luminosities of $\sim10^{36-37}$ \citep[e.g.,][]{sidoli09}. They seem to reside in their quiescent states ($L_{\mathrm{X}}\sim10^{32}~\lum$) only occasionally, and instead linger the majority of their time at levels of $\sim 10^{33-34}~\lum$ displaying X-ray spectra that are well fit by a powerlaw model with a photon index in the range 1--2 \citep[2--10 keV, e.g.,][]{sidoli08}. 
Slow (i.e., a few seconds to minutes) pulsations have been detected from a few of these systems \citep[e.g.,][]{sidoli09}. An HMXB configuration would be consistent with the possible detection of 5.25~s (0.19~Hz) pulsations from \xmmbron\ in \xmm\ observations \citep[][]{sakano05}. However, since the data analysis was limited by both statistics and exposure, the reliability of the coherent signal was considered highly uncertain by these authors and this result therefore needs to be verified.

\citet[][]{laylock05} report on $I$-band images of the field around \xmmbron, obtained during an X-ray outburst in 2005 early-June. Using the IMACS imaging spectrograph mounted at the {\it Magellan-Baade} telescope, these authors detect an optical source with $I=22.04\pm0.1$~mag and $R-I=1.95\pm0.2$, a low extinction ($A_{V}\sim7$~mag) and no variability. However, this object is located $\sim3\arcsec$ NE from the \chan\ coordinates of \xmmbron, lying outside the $1.5\arcsec$ positional uncertainty \citep[][]{wijnands06}, and is therefore likely not related. This implies that the optical counterpart of \xmmbron\ has a magnitude $I>25.6$~mag \citep[$3\sigma$ upper limit;][]{laylock05}. Finding an optical counterpart is hampered by the large extinction in the direction of the source. Using the relation of \citet{predehl1995}, a hydrogen column of $N_{\mathrm{H}}\sim 7.5 \times 10^{22}~\mathrm{cm}^{-2}$ (as inferred from fitting X-ray spectral data; see Section~\ref{subsec:xmm}) would translate into a visual extinction of $A_{\mathrm{V}}\sim42$ mag. Since the extinction is much lower at longer wavelengths, it might be more fruitful to search for a counterpart in the infrared. 

A recent study by \citet[][]{mauerhan2009} did not reveal any infrared objects associated with \xmmbron, up to a limiting magnitude of $K_s\lesssim15.6$~mag. The extinction in the $K_s$ band can be estimated as $A_{Ks}=0.062 \times A_V \sim 2.6$~mag \citep[][]{extinction2008}. Using the tables of \citet{drilling2000} and \citet{tokunaga2000} suggests that the survey of \citet[][]{mauerhan2009} should have enabled the detection of an O/B supergiant ($K_s \sim 11$~mag), as well as a main sequence star with spectral type earlier than B3V. Since most known HMXBs have donor stars with spectral types from O9V--B2V \citep[][]{negueruela1998}, \xmmbron\ is not likely to be an HMXB unless the source is more distant than 8 kpc. Its behavior therefore remains puzzling. 

Another possibility to explore is whether this source could be an LMXB in which a neutron star is accreting from the wind of an M-gaint companion. Currently, only 8 of such symbiotic X-ray binaries have been identified \citep[e.g.,][]{masetti2007,nespoli2010}. All of these systems show both long- and short-term X-ray variability and are characterized by 2--10 keV luminosities ranging between $\sim10^{32-35}~\lum$, although one object shows more intense X-ray emission of $L_{X}\sim10^{36-37}~\lum$ \citep[e.g.,][]{masetti2007}. However, for the extinction towards \xmmbron\ and a distance of 8 kpc, an M-type giant would have a magnitude of $K_s \sim 11-13$~mag \citep[][]{drilling2000,tokunaga2000}. The lack of a counterpart with $K_s \lesssim 15.6$~mag \citep[][]{mauerhan2009} therefore renders this scenario unlikely as well unless the source is located at a larger distance than 8 kpc.

\begin{table*}
\begin{threeparttable}[t]
\begin{center}
\caption[]{{Overview of the outburst properties and estimated (time-averaged) mass-accretion rates.}}
\begin{tabular}{l l l l l l l}
\hline
\hline
Source & Year &  $t_{\mathrm{ob}}$ & \~{F } & Duty cycle & $\langle \dot{M}\rangle _{\mathrm{ob}}$ & $\langle \dot{M} \rangle _{\mathrm{long}}$ \\
\hline
\ascabron\  &  &  &  & $10-30\%$ &  & $\sim (3-8)\times10^{-11}$ \\ 
& 2007--2008 & $>80$ & $>7 \times 10^{-3}$ & & $\sim3\times10^{-10}$ &  \\
 & 2006 & $>16$ &$\gtrsim 5 \times 10^{-4}$ & & $\sim1\times10^{-10}$ &  \\
\brontwee\  & &  & & $20-50\%$ &  & $\sim (7-14) \times 10^{-13}$ \\ 
& 2008 & $>8$ & $\gtrsim 7 \times 10^{-6}$ & & $\sim 3 \times10^{-12}$ &  \\
& 2006 & $>12$ & $\gtrsim 2 \times 10^{-5}$ & & $\sim 4 \times10^{-12}$ &  \\
\grsbron\  &  &  &  & $5-15\%$ &  & $\sim(2-6)\times 10^{-11}$ \\ 
& 2009 & $4-5$ & $\sim8\times10^{-4}$ & & $\sim5\times10^{-10}$ &  \\ 
& 2007 & $>13$ & $\gtrsim 1 \times 10^{-3}$ & & $\sim3\times10^{-10}$ &  \\
& 2006 & $<1$ & $\lesssim 3\times 10^{-6}$ & & $\sim8\times10^{-12}$ &  \\
\xmmbron\  &  &  &  & $5-50\%$ & & $\sim (1-10) \times 10^{-12}$ \\ 
& 2009 & $<2$ & $\lesssim3\times10^{-5}$& & $\sim4\times10^{-11}$ &  \\
& 2008 & $1-7$ & $\sim(2-10)\times10^{-5}$ & & $\sim6\times10^{-11}$ & \\
& 2008-low & $>16$ & $\gtrsim3\times10^{-5}$ & & $\sim2\times10^{-12}$ & \\
& 2007 & $<12$ & $\lesssim 4 \times 10^{-6}$ & & $\sim1\times10^{-12}$  &  \\
\bronacht\  & 2009 & $30-52$ & $\sim (9-20)\times10^{-5}$ & $5-20\%$ & $\sim1\times10^{-11}$ & $\sim(5-20)\times10^{-13}$ \\
\brondrie\ & 2006 & $2$ & $\sim 1\times10^{-5}$ & $1-5\%$ & $\sim2\times10^{-11}$ & $\sim(3-13)\times10^{-13}$ \\
\bronvier\ & 2006 & $2$ & $\sim 8\times10^{-6}$ & $\lesssim 5\%$ & $\sim2\times10^{-11}$ & $\lesssim 6\times10^{-13}$ \\
\bronvijf\ & 2006 & $5$ & $\sim 5\times10^{-6}$ & $\lesssim 14\%$ & $\sim3\times10^{-12}$ & $\lesssim 4\times10^{-13}$ \\
\hline
\end{tabular}
\label{tab:mdot}
\begin{tablenotes}
\item[]Note. -- The outburst duration, $t_{\mathrm{ob}}$, is expressed in weeks. \~{F} represents the fluency of the outburst in units of $\mathrm{erg~cm}^{-2}$ in the 2--10 keV energy band. $\langle \dot{M}\rangle _{\mathrm{ob}}$ is the estimated average accretion rate during outburst ($\mathrm{M}_{\odot}~\mathrm{yr}^{-1}$) assuming a neutron star primary with $M=1.4~\mathrm{M_{\odot}}$ and $R=10~\mathrm{km}$. The estimated time-averaged mass-accretion rate is given by $\langle \dot{M} \rangle _{\mathrm{long}}$. Note that only \ascabron\ and \grsbron\ are confirmed neutron star X-ray binaries, the other six sources have an unknown nature.
\end{tablenotes}
\end{center}
\end{threeparttable}
\end{table*}

\subsection{Mass-accretion rates}\label{subsec:mdot}
Amongst the different transients that are detected during the \swift/XRT monitoring observations of the GC there are two LMXBs (\ascabron\ and \grsbron), whereas the others remain unclassified. However, the energies involved in their outburst phenomena make it likely that these harbor accreting neutron stars or black holes. Recently, \citet{mauerhan2009} searched for near-infrared counterparts to X-ray sources located towards the GC. Their catalog reveals no counterparts for any of the unclassified transients detected during the \swift\ campaign, with a limiting magnitude of $K_s \lesssim15.6$~mag. This suggests that these systems are likely transient LMXBs rather than HMXBs \citep[][]{muno05_apj622,mauerhan2009}. It is therefore interesting to estimate the mean accretion rate during the outbursts of these systems, $\langle \dot{M}\rangle _{\mathrm{ob}}$, from which we can obtain an order of magnitude approximation for the long-term averaged mass-accretion rates, when combined with estimates of their duty cycles. 

The time-averaged mass-accretion rate, $\langle \dot{M} \rangle _{\mathrm{long}}$, is an important parameter for binary evolution models that attempt to explain the nature of low-luminosity LMXBs \citep[e.g.,][]{king_wijn06}. 
We refer to \citet[][]{degenaar09_gc} for the details of such calculations and the associated caveats. Here, we only calculate the (time-averaged) mass-accretion rates assuming a neutron star primary with $M_{\mathrm{NS}}=1.4~\mathrm{M_{\odot}}$ and $R_{\mathrm{NS}}=10$~km. These results are listed in Table~\ref{tab:mdot}. In case of a black hole accretor with $M_{\mathrm{BH}}=10~\mathrm{M_{\odot}}$ and $R_{\mathrm{BH}}=30$~km, the values given in this table can be multiplied with a factor $\sim0.4$, although one should bear in mind the caveats discussed in \citet[][]{degenaar09_gc}.

The duty cycles of \ascabron\ and \grsbron\ have been estimated in Sections~\ref{subsubsec:ascabron} and~\ref{subsubsec:grsbron}, respectively. For a discussion on the outburst and quiescent timescales of \brontwee\ and \xmmbron\ we refer to \citet[][]{degenaar09_gc}, since the new 2008--2009 data leaves those estimates unaltered. As mentioned in Section~\ref{sec:discuss}, the \swift\ monitoring observations detected activity of three transients in 2006 \citep[][]{degenaar09_gc}, that did not recur in 2008--2009. Since \bronvijf\ was in FOV only during a small number of pointings, we cannot refine the time-averaged mass-accretion rate for this source. However, for \brondrie\ and \bronvier, we can put further constraints on the time that these systems spend in quiescence. In 2008, daily observations were carried out for 36 consecutive weeks, which can thus be used as a lower limit on the quiescent timescale of the two transients. Both \brondrie\ and \bronvier\ exhibited an outburst with a duration of two weeks in 2006. This new constraint then puts their duty cycles at $\lesssim5\%$. Based on the 2006--2007 data set, duty cycles of $\lesssim6\%$ and $\lesssim8\%$ were estimated for \brondrie\ and \bronvier, respectively \citep[][]{degenaar09_gc}. Since the former was also active in 2003 \citep[][]{muno05_apj622}, the lower limit on its duty cycle is $\gtrsim1\%$ \citep[][]{degenaar09_gc}.

\bronacht\ was detected for the first time during the \swift/XRT observations performed in 2009. The source was observed with a peak outburst luminosity of $\sim2 \times 10^{35}~\lum$. Apart from the 2009 activity, \bronacht\ has been detected with a luminosity exceeding $1 \times 10^{34}~\lum$ only once before, in 1999 September with \chan\ \citep[][]{muno05_apj622}. This implies that the quiescent timescale of this source is less than 10~years. On the other hand, the source was not found active throughout the 2006--2008 \swift/XRT monitoring campaign. During those years, nearly daily observations were carried out, only interrupted for 17 weeks between 2006 November and 2007 March, for 15 weeks in the epoch 2007 November--2008 February (both due to Sun-angle constraints), and for 6 weeks between 2007 August 11--September 26 \citep[due to a safe-hold event;][]{swift_offline07}. If an outburst duration of 30--52~weeks is typical for this source, we can thus put a lower limit on the quiescent timescale of $\sim2.7$~years (139 weeks), although shorter outbursts might have been missed. The duty cycle of this source is then roughly between $\sim5-20\%$, which results in a time-averaged mass-accretion rate of $\langle \dot{M} \rangle _{\mathrm{long}}\sim(5-20)\times10^{-13}~\mdot$ (see Table~\ref{tab:mdot}). Despite the apparent long outburst duration, the estimated time-averaged accretion rate is amongst the lowest of the transients detected in the \swift/XRT monitoring campaign of the GC. 

It can be seen from Table~\ref{tab:mdot} that the two confirmed neutron star LMXBs \ascabron\ and \grsbron\ have estimated time-averaged mass-accretion rates of a few times $10^{-11}~\mdot$, which is not extraordinary low compared to other LMXBs. \xmmbron\ is also amongst the brightest transients detected during the \swift\ monitoring observations ($L_{\mathrm{X,peak}}\sim10^{36}~\lum$) and this system appears to recur quite often. This results in a relatively high time-averaged mass-accretion rate ($10^{-12}-10^{-11}~\mdot$) compared to the other five transients listed in Table~\ref{tab:mdot}, which have lower outburst luminosities and lower estimated rates of $\langle \dot{M} \rangle _{\mathrm{long}}\lesssim2\times10^{-12}~\mdot$. 

As mentioned above, the time-averaged mass-accretion rate is an important parameter for binary evolution models. \citet{king_wijn06} construct a theoretical toy model exploring the evolution of LMXBs at low accretion luminosities. The estimates of these authors show that if objects like \brontwee, \bronacht, \brondrie, \bronvier\ and \bronvijf\ are indeed X-ray binaries, their time-averaged mass-accretion rates suggests that the mass-donors are likely very low-mass or hydrogen-depleted stars. However, further refinement of their duty cycles and outburst energetics, as well as detailed evolutionary calculations, are required to grasp the nature of these peculiar objects.

\subsection{Summary of the campaign 2006--2009}\label{subsec:stat}
Starting in 2006 February and extending into 2009, the \swift/XRT monitoring campaign of the GC detected activity of 8 different transients in total, from which 14 distinct outbursts were observed. All sources have experienced outbursts with peak 2--10 keV luminosities $L_{\mathrm{X,peak}}\lesssim10^{36}~\lum$, although the two neutron star LMXBs \ascabron\ and \grsbron\ both displayed brighter outbursts as well ($L_{\mathrm{X,peak}}\sim10^{36-37}~\lum$). Two of the eight transients are newly discovered sources, which were both active in 2006 \citep[\bronvier\ and \bronvijf;][]{degenaar09_gc}. Four of the eight transients were observed to recur over the 4-year time span of this campaign (see Table~\ref{tab:mdot}) and have relatively short recurrence times. These transients show a different peak flux, duration and lightcurve morphology from outburst to outburst, which is also seen in brighter X-ray transients \citep[][]{chen97}.

Currently, there are $13$ X-ray transients exhibiting 2--10 keV peak luminosities $\gtrsim10^{34}~\lum$ known in the region covered by the \swift/XRT monitoring observations.\footnote{The \swift\ monitoring observations cover all sources listed in table A.1 of \citet{wijnands06} that are located within $\sim14'$ distance from \sgra. In addition to the 11 objects from their list, two new transients were discovered by \swift\ in 2006 \citep[\bronvier\ and \bronvijf;][]{degenaar09_gc}.} 
Out of these 13 transients, only 1A 1742--289 becomes brighter than $L_{\mathrm{X}}>10^{37}~\lum$ \citep[2--10 keV; see][and references therein]{wijnands06}, while the remaining 12 undergo low-luminosity outbursts. The three sources \ascabron, \grsbron\ and \xmmbron\ are the brightest amongst these and have 2--10 keV peak luminosities of $10^{36-37}~\lum$, but the other 9 transients have never been observed with luminosities exceeding $10^{36}~\lum$. From the 12 low-luminosity transients, 7 were observed to recur in the past decade and thus have relatively short recurrence times. The remaining 5 objects (\adcbron, \munotransient, \xmmbrontwee, \bronvier\ and \bronvijf) were seen active only once and thus seem to recur less often. 

Despite the fact that $>250$~ks of new \swift\ data was obtained, spread over almost daily observations in 2008 and 2009, no new transients were found. \citet{muno2009} suggested that given the extensive monitoring of the GC in the past years, all X-ray binaries that are located in that region and recur on a timescale of a decade have been identified by now. The galactic population of X-ray binaries (both LMXBs and HMXBs) is expected to encompass $\sim2000$ objects \citep[e.g.,][]{verbunt1995}. The region around \sgra\ that has been monitored by \chan, \xmm\ and \swift\ in the past decade covers $\sim1\%$ of the stellar mass in the galactic disk \citep[][]{pfahl2002}. In this region, $\sim20$ likely X-ray binaries have been identified \citep[][]{muno2009}. Most of these are transient sources and strikingly, the majority have very low 2--10 keV peak luminosities of $\lesssim10^{36}~\lum$ \citep[][]{muno05_apj622,wijnands06,degenaar09_gc}. 

The number of likely X-ray binaries that have been identified in the vicinity of \sgra\ is thus consistent with that expected from population synthesis models. However, the GC has been monitored with instruments sensitive enough to detect low-luminosity transients only in the past decade and several of the currently known systems appear to have relatively short recurrence times compared to brighter X-ray transients \citep[e.g.,][]{chen97}. Continued monitoring of the GC is therefore important to search for transient outbursts from new systems to better constrain the number of X-ray binaries located near \sgra, and to gain more insight into the duty cycles of known systems.

\section*{Acknowledgments}
We acknowledge the use of public data from the \textit{Swift} data archive. This work was supported by the Netherlands Organization for Scientific Research (NWO).

\bibliographystyle{aa}
\bibliography{15322}

\begin{thebibliography}{68}
\expandafter\ifx\csname natexlab\endcsname\relax\def\natexlab#1{#1}\fi

\bibitem[{{Arnaud}(1996)}]{xspec}
{Arnaud}, K.~A. 1996, in Astronomical Society of the Pacific Conference Series,
  Vol. 101, Astronomical Data Analysis Software and Systems V, ed. G.~H.
  {Jacoby} \& J.~{Barnes}, 17

\bibitem[{{Burrows} {et~al.}(2005){Burrows}, {Hill}, {Nousek}, {Kennea},
  {Wells}, {Osborne}, {Abbey}, {Beardmore}, {Mukerjee}, {Short}, {Chincarini},
  {Campana}, {Citterio}, {Moretti}, {Pagani}, {Tagliaferri}, {Giommi},
  {Capalbi}, {Tamburelli}, {Angelini}, {Cusumano}, {Br{\"a}uninger}, {Burkert},
  \& {Hartner}}]{burrows05}
{Burrows}, D.~N., {Hill}, J.~E., {Nousek}, J.~A., {et~al.} 2005, Space Science
  Reviews, 120, 165

\bibitem[{{Campana}(2009)}]{campana09}
{Campana}, S. 2009, \apj, 699, 1144

\bibitem[{{Cannizzo}(1993)}]{cannizzo1993}
{Cannizzo}, J.~K. 1993, \apj, 419, 318

\bibitem[{{Chelovekov} \& {Grebenev}(2007)}]{chelovekov07_ascabron}
{Chelovekov}, I.~V. \& {Grebenev}, S.~A. 2007, Astronomy Letters, 33, 807

\bibitem[{{Chen} {et~al.}(1997){Chen}, {Shrader}, \& {Livio}}]{chen97}
{Chen}, W., {Shrader}, C.~R., \& {Livio}, M. 1997, \apj, 491, 312

\bibitem[{{Chenevez} {et~al.}(2009){Chenevez}, {Kuulkers}, {Beckmann}, {Bird},
  {Brandt}, {Domingo}, {Ebisawa}, {Jonker}, {Kretschmar}, {Markwardt},
  {Oosterbroek}, {Paizis}, {Risquez}, {Sanchez-Fernandez}, {Shaw}, \&
  {Wijnands}}]{chenevez09}
{Chenevez}, J., {Kuulkers}, E., {Beckmann}, V., {et~al.} 2009, The Astronomer's
  Telegram, 2235

\bibitem[{{Cocchi} {et~al.}(1999){Cocchi}, {Bazzano}, {Natalucci}, {Ubertini},
  {Heise}, {Muller}, \& {in 't Zand}}]{cocchi99}
{Cocchi}, M., {Bazzano}, A., {Natalucci}, L., {et~al.} 1999, \aap, 346, L45

\bibitem[{{Cooper} \& {Narayan}(2007)}]{cooper07}
{Cooper}, R.~L. \& {Narayan}, R. 2007, \apj, 661, 468

\bibitem[{{Cornelisse} {et~al.}(2002){Cornelisse}, {Verbunt}, {in't Zand},
  {Kuulkers}, {Heise}, {Remillard}, {Cocchi}, {Natalucci}, {Bazzano}, \&
  {Ubertini}}]{cornelisse02}
{Cornelisse}, R., {Verbunt}, F., {in't Zand}, J.~J.~M., {et~al.} 2002, \aap,
  392, 885

\bibitem[{{Degenaar} \& {Wijnands}(2009)}]{degenaar09_gc}
{Degenaar}, N. \& {Wijnands}, R. 2009, \aap, 495, 547

\bibitem[{{Degenaar} {et~al.}(2010){Degenaar}, {Wijnands}, {Kennea}, \&
  {Gehrels}}]{degenaar2010_atel_asca}
{Degenaar}, N., {Wijnands}, R., {Kennea}, J., \& {Gehrels}, N. 2010, The
  Astronomer's Telegram, 2690

\bibitem[{{Degenaar} {et~al.}(2008{\natexlab{a}}){Degenaar}, {Wijnands}, \&
  {Muno}}]{deeg08_atel_gc}
{Degenaar}, N., {Wijnands}, R., \& {Muno}, M. 2008{\natexlab{a}}, The
  Astronomer's Telegram, 1531

\bibitem[{{Degenaar} {et~al.}(2008{\natexlab{b}}){Degenaar}, {Wijnands}, \&
  {Muno}}]{degenaar08_atel_gc_chan}
{Degenaar}, N., {Wijnands}, R., \& {Muno}, M. 2008{\natexlab{b}}, The
  Astronomer's Telegram, 1531

\bibitem[{{Del Santo} {et~al.}(2007){Del Santo}, {Sidoli}, {Mereghetti},
  {Bazzano}, {Tarana}, \& {Ubertini}}]{delsanto07}
{Del Santo}, M., {Sidoli}, L., {Mereghetti}, S., {et~al.} 2007, \aap, 468, L17

\bibitem[{{Drilling} \& {Landolt}(2000)}]{drilling2000}
{Drilling}, J.~S. \& {Landolt}, A.~U. 2000, {Normal Stars}, ed. {Cox, A.~N.},
  381--+

\bibitem[{{Gehrels}(2007)}]{swift_offline07}
{Gehrels}, N. 2007, GRB Coordinates Network, 6760

\bibitem[{{Heinke} {et~al.}(2009){Heinke}, {Cohn}, \& {Lugger}}]{heinke09_vfxt}
{Heinke}, C.~O., {Cohn}, H.~N., \& {Lugger}, P.~M. 2009, \apj, 692, 584

\bibitem[{{in 't Zand} {et~al.}(1991){in 't Zand}, {Heise}, {Brinkman},
  {Jager}, {Skinner}, {Patterson}, {Pan}, {Nottingham}, {Willmore}, \&
  {Al-Emam}}]{zand1991}
{in 't Zand}, J.~J.~M., {Heise}, J., {Brinkman}, A.~C., {et~al.} 1991, Advances
  in Space Research, 11, 187

\bibitem[{{in't Zand} {et~al.}(2005){in't Zand}, {Cumming}, {van der Sluys},
  {Verbunt}, \& {Pols}}]{zand05_ucxb}
{in't Zand}, J.~J.~M., {Cumming}, A., {van der Sluys}, M.~V., {Verbunt}, F., \&
  {Pols}, O.~R. 2005, \aap, 441, 675

\bibitem[{{in't Zand} {et~al.}(2007){in't Zand}, {Jonker}, \&
  {Markwardt}}]{zand07}
{in't Zand}, J.~J.~M., {Jonker}, P.~G., \& {Markwardt}, C.~B. 2007, \aap, 465,
  953

\bibitem[{{Kennea}(2009)}]{kennea09}
{Kennea}, J.~A. 2009, The Astronomer's Telegram, 2236

\bibitem[{{Kennea} {et~al.}(2006){Kennea}, {Burrows}, {Campana}, {Godet},
  {Nousek}, \& {Gehrels}}]{kennea06_atel753}
{Kennea}, J.~A., {Burrows}, D.~N., {Campana}, S., {et~al.} 2006, The
  Astronomer's Telegram, 753

\bibitem[{{Kennea} \& {The Swift/XRT team}(2006)}]{kennea_monit}
{Kennea}, J.~A. \& {The Swift/XRT team}. 2006, in Bulletin of the American
  Astronomical Society, Vol.~38, Bulletin of the American Astronomical Society,
  381

\bibitem[{{King} \& {Ritter}(1998)}]{king98}
{King}, A.~R. \& {Ritter}, H. 1998, \mnras, 293, L42

\bibitem[{{King} \& {Wijnands}(2006)}]{king_wijn06}
{King}, A.~R. \& {Wijnands}, R. 2006, \mnras, 366, L31

\bibitem[{{Kuulkers} {et~al.}(2007{\natexlab{a}}){Kuulkers}, {Shaw},
  {Chenevez}, {Brandt}, {Courvoisier}, {Domingo}, {Kretschmar}, {Markwardt},
  {Mowlavi}, {Paizis}, {Risquez}, {Sanchez-Fernandez}, \&
  {Wijnands}}]{kuulkers07_atel1005}
{Kuulkers}, E., {Shaw}, S., {Chenevez}, J., {et~al.} 2007{\natexlab{a}}, The
  Astronomer's Telegram, 1005

\bibitem[{{Kuulkers} {et~al.}(2007{\natexlab{b}}){Kuulkers}, {Shaw},
  {Chenevez}, {Brandt}, {Domingo}, {Kretschmar}, {Markwardt}, {Mowlavi},
  {Paizis}, {Risquez}, {Sanchez-Fernandez}, \&
  {Wijnands}}]{kuulkers07_atel1008}
{Kuulkers}, E., {Shaw}, S., {Chenevez}, J., {et~al.} 2007{\natexlab{b}}, The
  Astronomer's Telegram, 1008

\bibitem[{{Kuulkers} {et~al.}(2007{\natexlab{c}}){Kuulkers}, {Shaw}, {Paizis},
  {Chenevez}, {Brandt}, {Courvoisier}, {Domingo}, {Ebisawa}, {Kretschmar},
  {Markwardt}, {Mowlavi}, {Oosterbroek}, {Orr}, {R{\'{\i}}squez},
  {Sanchez-Fernandez}, \& {Wijnands}}]{kuulkers07}
{Kuulkers}, E., {Shaw}, S.~E., {Paizis}, A., {et~al.} 2007{\natexlab{c}}, \aap,
  466, 595

\bibitem[{{Lasota}(2001)}]{lasota01}
{Lasota}, J.-P. 2001, New Astronomy Review, 45, 449

\bibitem[{{Lasota}(2007)}]{lasota07}
{Lasota}, J.-P. 2007, Comptes Rendus Physique, 8, 45

\bibitem[{{Laycock} {et~al.}(2005){Laycock}, {Zhao}, {Torres}, {Wijnands},
  {Steeghs}, {Grindlay}, {Hong}, \& {Jonker}}]{laylock05}
{Laycock}, S., {Zhao}, P., {Torres}, M.~A.~P., {et~al.} 2005, The Astronomer's
  Telegram, 522

\bibitem[{{Maeda} {et~al.}(1996){Maeda}, {Koyama}, {Sakano}, {Takeshima}, \&
  {Yamauchi}}]{maeda1996}
{Maeda}, Y., {Koyama}, K., {Sakano}, M., {Takeshima}, T., \& {Yamauchi}, S.
  1996, \pasj, 48, 417

\bibitem[{{Masetti} {et~al.}(2007){Masetti}, {Landi}, {Pretorius}, {Sguera},
  {Bird}, {Perri}, {Charles}, {Kennea}, {Malizia}, \& {Ubertini}}]{masetti2007}
{Masetti}, N., {Landi}, R., {Pretorius}, M.~L., {et~al.} 2007, \aap, 470, 331

\bibitem[{{Mauerhan} {et~al.}(2009){Mauerhan}, {Muno}, {Morris}, {Bauer},
  {Nishiyama}, \& {Nagata}}]{mauerhan2009}
{Mauerhan}, J.~C., {Muno}, M.~P., {Morris}, M.~R., {et~al.} 2009, \apj, 703, 30

\bibitem[{{Menou} {et~al.}(1999){Menou}, {Esin}, {Narayan}, {Garcia}, {Lasota},
  \& {McClintock}}]{menou99}
{Menou}, K., {Esin}, A.~A., {Narayan}, R., {et~al.} 1999, \apj, 520, 276

\bibitem[{{Muno} {et~al.}(2004){Muno}, {Arabadjis}, {Baganoff}, {Bautz},
  {Brandt}, {Broos}, {Feigelson}, {Garmire}, {Morris}, \&
  {Ricker}}]{muno04_apj613}
{Muno}, M.~P., {Arabadjis}, J.~S., {Baganoff}, F.~K., {et~al.} 2004, \apj, 613,
  1179

\bibitem[{{Muno} {et~al.}(2003{\natexlab{a}}){Muno}, {Baganoff}, \&
  {Arabadjis}}]{muno03_grs}
{Muno}, M.~P., {Baganoff}, F.~K., \& {Arabadjis}, J.~S. 2003{\natexlab{a}},
  \apj, 598, 474

\bibitem[{{Muno} {et~al.}(2003{\natexlab{b}}){Muno}, {Baganoff}, {Bautz},
  {Brandt}, {Broos}, {Feigelson}, {Garmire}, {Morris}, {Ricker}, \&
  {Townsley}}]{muno03}
{Muno}, M.~P., {Baganoff}, F.~K., {Bautz}, M.~W., {et~al.} 2003{\natexlab{b}},
  \apj, 589, 225

\bibitem[{{Muno} {et~al.}(2009){Muno}, {Bauer}, {Baganoff}, {Bandyopadhyay},
  {Bower}, {Brandt}, {Broos}, {Cotera}, {Eikenberry}, {Garmire}, {Hyman},
  {Kassim}, {Lang}, {Lazio}, {Law}, {Mauerhan}, {Morris}, {Nagata},
  {Nishiyama}, {Park}, {Ram{\`i}rez}, {Stolovy}, {Wijnands}, {Wang}, {Wang}, \&
  {Yusef-Zadeh}}]{muno2009}
{Muno}, M.~P., {Bauer}, F.~E., {Baganoff}, F.~K., {et~al.} 2009, \apjs, 181,
  110

\bibitem[{{Muno} {et~al.}(2005){Muno}, {Pfahl}, {Baganoff}, {Brandt}, {Ghez},
  {Lu}, \& {Morris}}]{muno05_apj622}
{Muno}, M.~P., {Pfahl}, E., {Baganoff}, F.~K., {et~al.} 2005, \apjl, 622, L113

\bibitem[{{Muno} {et~al.}(2007){Muno}, {Wijnands}, {Wang}, {Park}, {Brandt},
  {Bauer}, \& {Wang}}]{muno07_atel1013}
{Muno}, M.~P., {Wijnands}, R., {Wang}, Q.~D., {et~al.} 2007, The Astronomer's
  Telegram, 1013

\bibitem[{{Narayan} {et~al.}(1997){Narayan}, {Garcia}, \&
  {McClintock}}]{narayan97}
{Narayan}, R., {Garcia}, M.~R., \& {McClintock}, J.~E. 1997, \apjl, 478, L79+

\bibitem[{{Negueruela}(1998)}]{negueruela1998}
{Negueruela}, I. 1998, \aap, 338, 505

\bibitem[{{Negueruela}(2004)}]{negueruela04}
{Negueruela}, I. 2004, in Revista Mexicana de Astronomia y Astrofisica
  Conference Series, Vol.~20, Revista Mexicana de Astronomia y Astrofisica
  Conference Series, ed. {G.~Tovmassian \& E.~Sion}, 55--56

\bibitem[{{Negueruela} {et~al.}(2006){Negueruela}, {Smith}, {Reig}, {Chaty}, \&
  {Torrej{\'o}n}}]{negueruela06}
{Negueruela}, I., {Smith}, D.~M., {Reig}, P., {Chaty}, S., \& {Torrej{\'o}n},
  J.~M. 2006, in ESA Special Publication, Vol. 604, The X-ray Universe 2005,
  ed. {A.~Wilson}, 165--+

\bibitem[{{Nespoli} {et~al.}(2010){Nespoli}, {Fabregat}, \&
  {Mennickent}}]{nespoli2010}
{Nespoli}, E., {Fabregat}, J., \& {Mennickent}, R.~E. 2010, \aap, 516, A94+

\bibitem[{{Nishiyama} {et~al.}(2008){Nishiyama}, {Nagata}, {Tamura}, {Kandori},
  {Hatano}, {Sato}, \& {Sugitani}}]{extinction2008}
{Nishiyama}, S., {Nagata}, T., {Tamura}, M., {et~al.} 2008, \apj, 680, 1174

\bibitem[{{Peng} {et~al.}(2007){Peng}, {Brown}, \& {Truran}}]{peng2007}
{Peng}, F., {Brown}, E.~F., \& {Truran}, J.~W. 2007, \apj, 654, 1022

\bibitem[{{Pfahl} {et~al.}(2002){Pfahl}, {Rappaport}, \&
  {Podsiadlowski}}]{pfahl2002}
{Pfahl}, E., {Rappaport}, S., \& {Podsiadlowski}, P. 2002, \apjl, 571, L37

\bibitem[{{Ponti} {et~al.}(2009){Ponti}, {Trap}, {Goldwurm}, {Ferrando},
  {Terrier}, {Belanger}, {Genzel}, {Gillessen}, {Hasinger}, {Dodds-Eden},
  {Predehl}, {Aschenbach}, {Porquet}, {Grosso}, {Clenet}, {Rouan}, {Sakano},
  {Warwick}, {Melia}, {Farhad}, \& {Reid}}]{ponti09}
{Ponti}, G., {Trap}, G., {Goldwurm}, A., {et~al.} 2009, The Astronomer's
  Telegram, 2038

\bibitem[{{Porquet} {et~al.}(2005){Porquet}, {Grosso}, {Burwitz}, {Andronov},
  {Aschenbach}, {Predehl}, \& {Warwick}}]{porquet05}
{Porquet}, D., {Grosso}, N., {Burwitz}, V., {et~al.} 2005, \aap, 430, L9

\bibitem[{{Porquet} {et~al.}(2007){Porquet}, {Grosso}, {Goldwurm}, {Sakano},
  {Belanger}, {Ferrando}, {Hasinger}, {Aschenbach}, {Predhel}, {Tanaka},
  {Genzel}, {Yusef-Zadeh}, {Warwick}, \& {Melia}}]{porquet07}
{Porquet}, D., {Grosso}, N., {Goldwurm}, A., {et~al.} 2007, The Astronomer's
  Telegram, 1058

\bibitem[{{Predehl} \& {Schmitt}(1995)}]{predehl1995}
{Predehl}, P. \& {Schmitt}, J.~H.~M.~M. 1995, \aap, 293, 889

\bibitem[{{Sakano} {et~al.}(2002){Sakano}, {Koyama}, {Murakami}, {Maeda}, \&
  {Yamauchi}}]{sakano02}
{Sakano}, M., {Koyama}, K., {Murakami}, H., {Maeda}, Y., \& {Yamauchi}, S.
  2002, \apjs, 138, 19

\bibitem[{{Sakano} {et~al.}(2005){Sakano}, {Warwick}, {Decourchelle}, \&
  {Wang}}]{sakano05}
{Sakano}, M., {Warwick}, R.~S., {Decourchelle}, A., \& {Wang}, Q.~D. 2005,
  \mnras, 357, 1211

\bibitem[{{Sidoli}(2009)}]{sidoli09}
{Sidoli}, L. 2009, Advances in Space Research, 43, 1464

\bibitem[{{Sidoli} {et~al.}(1999){Sidoli}, {Mereghetti}, {Israel},
  {Chiappetti}, {Treves}, \& {Orlandini}}]{sidoli99}
{Sidoli}, L., {Mereghetti}, S., {Israel}, G.~L., {et~al.} 1999, \apj, 525, 215

\bibitem[{{Sidoli} {et~al.}(2008){Sidoli}, {Romano}, {Mangano}, {Pellizzoni},
  {Kennea}, {Cusumano}, {Vercellone}, {Paizis}, {Burrows}, \&
  {Gehrels}}]{sidoli08}
{Sidoli}, L., {Romano}, P., {Mangano}, V., {et~al.} 2008, \apj, 687, 1230

\bibitem[{{Sunyaev}(1990)}]{sunyaev1990}
{Sunyaev}, R. 1990, \iaucirc, 5104

\bibitem[{{Tokunaga}(2000)}]{tokunaga2000}
{Tokunaga}, A.~T. 2000, {Infrared Astronomy}, ed. {Cox, A.~N.}, 143--+

\bibitem[{{Trap} {et~al.}(2009){Trap}, {Falanga}, {Goldwurm}, {Bozzo},
  {Terrier}, {Ferrando}, {Porquet}, {Grosso}, \& {Sakano}}]{trap09}
{Trap}, G., {Falanga}, M., {Goldwurm}, A., {et~al.} 2009, \aap, 504, 501

\bibitem[{{Verbunt} \& {van den Heuvel}(1995)}]{verbunt1995}
{Verbunt}, F. \& {van den Heuvel}, E.~P.~J. 1995, in X-ray binaries, p. 457 -
  494, ed. {W.~H.~G.~Lewin, J.~van Paradijs, \& E.~P.~J.~van den Heuvel},
  457--494

\bibitem[{{Wijnands} {et~al.}(2006{\natexlab{a}}){Wijnands}, {in't Zand},
  {Rupen}, {Maccarone}, {Homan}, {Cornelisse}, {Fender}, {Grindlay}, {van der
  Klis}, {Kuulkers}, {Markwardt}, {Miller-Jones}, \& {Wang}}]{wijnands06}
{Wijnands}, R., {in't Zand}, J.~J.~M., {Rupen}, M., {et~al.}
  2006{\natexlab{a}}, \aap, 449, 1117

\bibitem[{{Wijnands} {et~al.}(2007){Wijnands}, {Klein-Wolt}, {Kuulkers},
  {Shaw}, {Chenevez}, {Brandt}, {Courvoisier}, {Domingo}, {Kretschmar},
  {Markwardt}, {Mowlavi}, {Paizis}, {Risquez}, \&
  {Sanchez-Fernandez}}]{wijnands07_atel1006}
{Wijnands}, R., {Klein-Wolt}, M., {Kuulkers}, E., {et~al.} 2007, The
  Astronomer's Telegram, 1006

\bibitem[{{Wijnands} {et~al.}(2006{\natexlab{b}}){Wijnands}, {Kuulkers},
  {Muno}, {Cackett}, {in't Zand}, {Maccarone}, {Fender}, {Grindlay}, {Homan},
  {Rupen}, {Cornelisse}, {Miller-Jones}, {van der Klis}, {Markwardt}, \&
  {Wang}}]{wijnands06_atel892}
{Wijnands}, R., {Kuulkers}, E., {Muno}, M., {et~al.} 2006{\natexlab{b}}, The
  Astronomer's Telegram, 892

\bibitem[{{Wijnands} {et~al.}(2009){Wijnands}, {Rol}, {Cackett}, {Starling}, \&
  {Remillard}}]{wijnands09}
{Wijnands}, R., {Rol}, E., {Cackett}, E., {Starling}, R.~L.~C., \& {Remillard},
  R.~A. 2009, \mnras, 393, 126

\bibitem[{{Wijnands} {et~al.}(2005){Wijnands}, {Rupen}, {Steeghs}, {Wang},
  {Cackett}, {Kuulkers}, {Fender}, {Cornelisse}, {Maccarone}, {Grindlay},
  {Miller-Jones}, {Klis}, {Homan}, {in't Zand}, \&
  {Markwardt}}]{wijnands05_atel638}
{Wijnands}, R., {Rupen}, M., {Steeghs}, D., {et~al.} 2005, The Astronomer's
  Telegram, 638

\end{thebibliography}

\end{document}